%% file: Main.tex
\documentclass[journal]{IEEEtran}

\usepackage{cite} % Sort and compress reference ranges.
\usepackage{booktabs}
\usepackage{threeparttable}

\usepackage[dvipsnames]{xcolor}
\usepackage{adjustbox}
\usepackage{graphicx}
\usepackage[caption=false,font=footnotesize]{subfig}

\usepackage{dblfloatfix}
\usepackage{comment}
\usepackage{gensymb}

\usepackage{eqparbox}
\usepackage{algorithm}
\usepackage[noend]{algorithmic}

\usepackage{amsmath,amsfonts,amssymb,amsthm}

\DeclareMathOperator*{\argmax}{arg\,max}
\DeclareMathOperator*{\mean}{mean}
\usepackage{mathtools}
\usepackage{mathrsfs}	%	mathscr
\usepackage{subdepth}	%	uniform subscript position

\usepackage{array}	%	better appearance for array and tabular environments
\usepackage[inline]{enumitem} % enumerate stuff in line

\usepackage{hyperref}
\usepackage[sort,compress,capitalise]{cleveref}	%	...
\usepackage{url}

\usepackage{xcolor} % Coloured text
\usepackage{xspace} % Intelligent spacing (acronym commands) 
\usepackage{lipsum}

\usepackage{tikz}
\usetikzlibrary{babel}  % Avoid problems due to babel library
\usepackage[straightvoltages,nooldvoltagedirection]{circuitikz}
\usetikzlibrary{shapes,arrows}
\usepackage{verbatim}
\usepackage{multirow}

\begin{document}

\title{Functional Basis Analysis for the Characterization of Power System Signal Dynamics: \\
Formulation, Implementation and Validation}

\author{Alexandra Karpilow,~\IEEEmembership{Student Member,~IEEE,}
        Asja Dervi\v{s}kadi\'{c},~\IEEEmembership{Member,~IEEE,}\\
        Mario Paolone,~\IEEEmembership{Senior Member,~IEEE}% <-this % stops 
        \thanks{A. Karpilow and M. Paolone are with the École Polytechnique Fédérale de Lausanne (EPFL), Switzerland. (e-mail: alexandra.karpilow@epfl.ch, mario.paolone@epfl.ch). A. Dervi\v{s}kadi\'{c} is with Swissgrid AG, Aarau, Switzerland. (e-mail:Asja.Derviskadic@swissgrid.ch) }
}
\maketitle

\begin{abstract}
With the integration of distributed energy resources and the trend towards low-inertia power grids, the frequency and severity of grid dynamics is expected to increase. Conventional phasor-based signal processing methods are proving to be insufficient in the analysis of non-stationary AC voltage and current waveforms, while the computational complexity of many dynamic signal analysis techniques hinders their deployment in operational embedded systems. This paper presents the Functional Basis Analysis (FBA), a signal processing tool capable of capturing the full broadband nature of signal dynamics in power grids while maintaining a streamlined design for real-time monitoring applications. Relying on the Hilbert transform and optimization techniques, the FBA can be user-engineered to identify and characterize combinations of several of the most common signal dynamics in power grids, including amplitude/phase modulations, frequency ramps and steps. This paper describes the theoretical basis and design of the FBA as well as the 
deployment of the algorithm in embedded hardware systems, with adaptations made to consider latency requirements, finite memory capacity, and fixed-point precision arithmetic. For validation, a PMU calibrator is used to evaluate and compare the algorithm's performance to state-of-the-art static and dynamic phasor methods. The test outcomes demonstrate the FBA method's suitability for implementation in embedded systems to improve grid situational awareness during severe grid events.
\end{abstract}

\input{Chapters/1_Introduction}
\input{Chapters/2_Algorithm}

\input{Chapters/3_LVImplementation}

\input{Chapters/4_Experiments}

\input{Chapters/5_Results}

\input{Chapters/6_Conclusions}

\input{Chapters/Appendix}

\IEEEpeerreviewmaketitle
\bibliographystyle{IEEEtran}
\bibliography{ref.bib}
\end{document}

%% file: Chapters/1_Introduction.tex
\section{Introduction}

As traditional power plants are phased out and replaced with converter-interfaced distributed energy resources, next-generation power grids are expected to experience heightened and more frequent dynamics, due to the reduced system inertia, the presence of power electronics and nonlinear loads, and the intermittency of renewable energy sources \cite{NREL_inertia}. Recently recorded events in Australia \cite{AustraliaBlackout}, Europe \cite{entsoe2017-EUoscillation}, and the U.S.A \cite{CA-2017}, demonstrate the trend towards more severe dynamics in a power grid governed less and less by the stabilizing effects of synchronous generators.
In light of this evolution, and the subsequent need for reliable monitoring technologies, Phasor Measurement Units (PMUs) have emerged as an invaluable tool for situational awareness, wide-area protection and control schemes. Capable of providing distributed and synchronized measurements at high reporting rates, PMUs can improve state estimation, voltage stability analysis, oscillation detection, protection and fault location \cite{EPRI_PMU}, with applications in both high voltage transmission networks and power distribution systems.

Most PMUs rely on the static phasor model and the Discrete Fourier Transform (DFT) due to its simple implementation, low latency and short window length, and capacity to reduce the impact of harmonics and spectral leakage \cite{iIPDFT}. However, the static phasor model, as a narrowband representation at a single frequency, is inherently incapable of capturing the broadband spectral nature of signal dynamics common in modern power grids. Recent literature, investigating the appropriateness of applying phasor-based models in the analysis of transient waveforms \cite{Roadmap_CPOW,Kirkham_NonStationarySignals, phasor_transient_conditions, RedefineFreq_PhaseSteps}, finds that static phasor estimation during extreme grid events can lead to misinterpretation of the grid state and incorrect operator actions \cite{FieldMeas, BeyondPhasors}.

Alternatively, the dynamic phasor model can account for variations of the amplitude and phase within the observation window \cite{ DynPhasor_Serna}. Common algorithms in this class use Taylor series expansion or the Taylor-Fourier transform, combined with Weighted Least Squares or Compressed Sensing theory, to approximate the first and second derivatives to improve the estimation of an underlying fundamental tone \cite{DynPhasor_Serna,Castello_TaylorFourier,CompressiveSensing_TF_Frigo,TaylorWLS_dynamicphasor}. However, these methods are often computationally expensive, precluding them from being implemented in standard hardware platforms, and can still provide erroneous results for sudden transients like phase and amplitude steps \cite{PaoloBook}. Several time-domain fitting techniques have also been proposed. In \cite{Kalman}, the authors employ a Kalman filter and compare the predicted waveform to the true input to detect sudden transients. However, Kalman filters are notoriously sensitive to the tuning of initialization parameters and estimation of the process noise covariance matrix \cite{ProblemKalman}. In \cite{AdaptiveWindowingWavelet}, wavelet analysis is used to first identify the location of discontinuities in the waveform, followed by an adaptive windowing technique to fit a quadratic polynomial signal model to pre- and post-event data. Separately, the recursive least squares (RLS) technique is capable of phasor estimation in the presence of decaying DC offsets and frequency variations. In \cite{DRLS}, a decoupled RLS variation is proposed to recursively update the inverse estimation matrix and reduce the computational complexity, which has hindered the practical implementation of RLS in embedded systems. However, the study focuses on the extraction of the static phasor rather than the characterization of the signal dynamic. 

While PMUs perform "lossy compression" (i.e., the signal dynamics present in the waveform can not be recovered from the compressed phasor model), an alternative strategy is to directly stream high resolution continuous point-on-wave (CPOW) data \cite{Roadmap_CPOW}, capturing the full broadband nature of the signal. While CPOW requires high data bandwidth and storage, often limiting it to local analysis or event-triggered streaming, recent studies are investigating ways to efficiently encode and compress measurement samples \cite{CPOW}.
Although CPOW is beneficial for post-processing, the inherent analysis delay renders it unsuitable for time-critical applications. To summarize, the major challenges in signal analysis of power system dynamics include proper model selection, resilience to or identification of dynamics, and implementation limitations (e.g., latency, computational complexity).

Contributing to this field of study, this paper proposes a method, titled the Functional Basis Analysis (FBA), to compress signal dynamic waveforms with better accuracy and flexibility than a PMU and reduced data streaming and storage requirements, relative to the CPOW approach. The preliminary concept of the FBA method was introduced in a first seminal paper \cite{BeyondPhasors} and developed further by the authors in \cite{PwrTech}, \cite{SGSMA} and \cite{PSCC}. By exploiting properties of complex analytic signals, constructed via the Hilbert transform (HT), the algorithms apply dictionaries of parameterized models to identify signal dynamics common to power systems. However, previous versions of the FBA required long observation windows, ideal Hilbert filters, or applied limiting assumptions to the signal model, constraining the set of characterizable signals. Furthermore, the algorithms were computationally burdensome, requiring storage of large look-up tables and scaling poorly with expanding dynamic parameter ranges.

In response to these challenges, this paper introduces a novel FBA formulation that assumes a flexible model and is tailored specifically for deployment in embedded devices to enable real-time signal analysis. 
The original contributions of this paper includes: 1) computationally efficient algorithms for the characterization of multidynamic signals, 2) implementation of the FBA method on an embedded device, and 3) experimental validation of the FBA prototype using a PMU calibrator.

The paper is structured as follows: Section \ref{sec:theory} introduces the signal dynamic models and basics of analytic signals, while Section \ref{sec:proposed_alg} outlines the FBA algorithm. Section \ref{sec:LVimplementation} dives into the hardware implementation and associated computational costs. Next, Section \ref{sec:Experimental_Validation} details the experimental setup and evaluates the FBA's performance against static and dynamic phasor estimation algorithms.
Finally, Section \ref{sec:Conclusion} summarizes the findings and recommends future applications of this work.

%% file: Chapters/2_Algorithm.tex
\section{Analytic Signals and Signal Dynamics} \label{sec:theory}%\label{sec:theory}
The analytic signal is defined as the complex waveform $\Tilde{x}(t)=x(t)+j x_{Im}(t)$. The imaginary component relates to the real component via the Hilbert transform $\mathcal{H}$, a linear operator that, for a generic time-varying signal $x(t)$, is defined as \cite{Hilbert}:
\begin{equation}\label{eqn_HT}
    x_{Im}(t)=\mathcal{H}[x(t)]=\frac{1}{\pi} \int_{-\infty}^{+\infty}\frac{x(\tau)}{t-\tau}d\tau= \frac{1}{\pi t}*x(t)
\end{equation}
where $*$ indicates convolution. In the frequency domain, this equates to a linear phase shift filter:
\begin{equation}\label{eqn_HT_freq}
    \mathcal{F}[\mathcal{H}(x(t))]=\mathcal{F}[\frac{1}{\pi t}]\mathcal{F}[x(t)]=-j sgn(\omega) X(\omega)
\end{equation}

where $\mathcal{F}$ is the Fourier transform and the multiplier $-j=e^{-j \pi/2}$ is the equivalent of a $90 \degree$ phase shift. Consequently, the
complex analytic signal has a one-sided spectrum as the negative frequency image is canceled out \cite{HilbertFilter_Summary,HilbertFilter_guide}. Furthermore, the instantaneous envelope, phase and frequency can be extracted from the complex time-domain signal \cite{Hilbert}.

 For power systems, signal dynamics can be generically modeled with the following discretized function \cite{def_ref_calibration}:
 \begin{equation} \label{eq:general model}
    x(t_l)=A_0(1+g_A(t_l))cos(2\pi f_0 t_l + g_\phi (t_l) 
    +\varphi_0)
\end{equation}
where $t_l=lT_s$, for $l=0...(L-1)$, are a finite set of timestamps for a given sampling frequency $F_s=1/T_s$ and number of samples $L$. $A_0$, $f_0$, and $\varphi_0$ are the fundamental amplitude, frequency and phase, respectively, and $g_A(t_l)$ and $g_\phi(t_l)$ represent the variation in amplitude and argument. The analytic signal \footnote{\eqref{eq:analyticform} is derived using the property that the HT of the product of two signals with non-overlapping spectra is equal to the product of the low-frequency term and the HT of the high-frequency term. \textcolor{black}{For justification see} \cite{Hilbert}.} of \eqref{eq:general model} is then:
\begin{equation}
    \Tilde{x}(t_l)=A_0(1+g_A(t_l))e^{{\textit{j}}(2\pi f_0 t_l + g_\phi (t_l)+\varphi_0}). \label{eq:analyticform}
\end{equation}
This paper analyzes several common signal dynamics in power grids, including amplitude modulations (AM), phase modulations (PM), amplitude steps (AS), phase steps (PS), and linear frequency ramps (FR), using the models outlined in Table \ref{tab:model}. Note that $h(t_l)$ is the unit step function, $(a_\text{AM}, a_\text{PM})$, $(\varphi_\text{AM}, \varphi_\text{PM})$, and $(f_\text{AM}, f_\text{PM})$, are the modulating magnitude, phase and frequency for AM and PM, respectively, $(a_\text{AS},a_\text{PS})$ and $(t_\text{AS},t_\text{PS})$ are the step magnitude and location for AS and PS, respectively, and $R$ is the ramp rate.

\begin{table}[]
    \centering
        \caption{Signal dynamic models.} \label{tab:model}

    \begin{tabular}{ll}
    \toprule
        \textbf{Signal Dynamic} & \textbf{Model} \\ 
        \midrule
        Amplitude Modulations (AM) & $g_A(t_l)=a_\text{AM} sin(2\pi f_\text{AM} t_l+\varphi_\text{AM})$  \\ 
        Phase Modulations (PM) & $g_\phi(t_l)=a_\text{PM} sin(2\pi f_\text{PM} t_l+\varphi_\text{PM})$  \\ 
        Amplitude Steps (AS) & $g_A(t_l)=a_\text{AS} h(t_l-t_\text{AS})$  \\ 
        Phase Steps (PS) &  $g_\phi(t_l)=a_\text{PS} h(t_l-t_\text{PS})$  \\ 
        Frequency Ramps (FR) & $g_\phi(t_l)=R \pi t_l^2$ \\ \bottomrule
    \end{tabular}
\end{table}

\section{Functional Basis Analysis}\label{sec:proposed_alg}

In the field of signal processing and Compressed Sensing, signal decomposition is a fundamental operation to extract meaningful sub-components and underlying structures from measured data \cite{SignalProcessingBook}. The FBA method relies heavily on this core theory in two ways: first, the complex analytic signal is separated into the instantaneous argument and amplitude for independent analysis. Secondly, the FBA uses several functional bases, specifically designed to capture signal dynamics expected in power systems (e.g., AM, FR, etc.), and extracts the expansion coefficients that best represent the input waveform with respect to these bases.

\subsection{Envelope Analysis}
The envelope of the analytic signal is modeled as:
\begin{equation}
    x_A(t_l)=|\Tilde{x}(t_l)|=A_0(1+g_A(t_l))
\end{equation}
and is assumed to be an AM, AS or  steady state (SS). If no step is detected in the window (see Section \ref{sec:AS}) the envelope is approximated by an AM. In the case of SS signals, the estimated magnitude or frequency is very small and the reconstructed envelope is almost constant.

\subsubsection{Amplitude Modulations}\label{sec:AM}
The parameters of an AM are challenging to fit as less than a full cycle of low frequency modulations are present in the window. As an alternative to DFT analysis \cite{PSCC}, which suffers from spectral leakage since the modulations are close to DC,  a time domain four-parameter fitting method similar to \cite{4paramfit} is proposed.

The envelope is modeled as an amplitude modulation:
\begin{equation}\label{eq:model_am}
    x_\text{AM}(t_l)=\gamma_0+\gamma_1 sin(2\pi f_\text{AM} t_l) + \gamma_2 cos(2\pi f_\text{AM} t_l),
\end{equation}
where $A_0 a_\text{AM}=\sqrt{\gamma_1^2+\gamma_2^2}$, $A_0=\gamma_0$, and $\varphi_\text{AM}=atan(\frac{\gamma_2}{\gamma_1})$ if $\gamma_1>0$ and $\varphi_\text{AM}=atan(\frac{\gamma_2}{\gamma_1})+\pi$ if $\gamma_1<0$ (see Table \ref{tab:model}). %if gamma_1 is less than zero then varphi +pi
Given $f_\text{AM}$, \eqref{eq:model_am} is linear relative to $\boldsymbol{\gamma}$ and can be solved with the least squares (LS) estimator. The objective function is:
\begin{equation}
   \underset{\boldsymbol{\gamma}}{\arg\min} \lVert \textbf{D}_\text{AM}(f_\text{AM})\boldsymbol{\gamma}-\textbf{x}_A\rVert_2
\end{equation} 
where $\boldsymbol{\gamma}=\begin{bmatrix}
\gamma_0, \gamma_1,\gamma_2
\end{bmatrix}^T, \textbf{x}_A=\begin{bmatrix}
x_A(t_0)...x_A(t_{L-1})
\end{bmatrix}^T$ and

\begin{equation}
\begin{split}
\textbf{D}_\text{AM}&(f_\text{AM}) \\ 
&=\begin{bmatrix}
    1 & sin(2\pi f_\text{AM} t_0) &  cos(2\pi f_\text{AM} t_0)\\
    \vdots & \vdots & \vdots \\
     1 & sin(2\pi f_\text{AM} t_{L-1}) & cos(2\pi f_\text{AM} t_{L-1}) 
    \end{bmatrix}
    \end{split}
\end{equation}
\\
Reformulating this problem in terms of signal decomposition theory: the input vector $\textbf{x}_A$ is projected onto the subspace spanned by the columns of the basis matrix $\textbf{D}_\text{AM}(f_\text{AM})$ to find the expansion coefficients $\boldsymbol{\gamma}$ \cite{SignalProcessing_book}. The unique 3-parameter solution is the closest approximation of the input in the subspace (i.e., the best approximation for the given model):
\begin{equation} \label{eq:AM_sol}
    \boldsymbol{\gamma}^*(f_\text{AM})=\boldsymbol{\Lambda}_\text{AM}^{-1}(f_\text{AM})\textbf{D}_\text{AM}(f_\text{AM})^T\textbf{x}_A
\end{equation}
where $\boldsymbol{\Lambda}_\text{AM}(f_\text{AM})=\textbf{D}_\text{AM}(f_\text{AM})^T\textbf{D}_\text{AM}(f_\text{AM})$ is the Gram matrix. To estimate $f_\text{AM}$, the Golden Section Search (GSS) \cite{GSS} is applied to iteratively approach the solution within a fixed range $[f_{lb}, f_{ub}]$ by minimizing the residual:
\begin{equation}\label{eq:AM_res}
    \rho_\text{AM}(f_\text{AM})=\sum_{l=0}^{L-1}(x_\text{AM}(t_l,f_\text{AM},\boldsymbol{\gamma}(f_\text{AM}))-x_A(t_l))^2
\end{equation}
where $x_\text{AM}(t_l)$ is the reconstructed envelope in \eqref{eq:model_am} for a given $f_\text{AM}$ and a parameter set $\boldsymbol{\gamma}(f_\text{AM})$, computed from \eqref{eq:AM_sol}. 

At each iteration, the GSS algorithm evaluates and compares the residual at four frequencies $f_{lb}<f_1<f_2<f_{ub}$ within the current interval, in order to determine the contracted interval of the next iteration (i.e., the "direction" of the minimum). However, while \ref{eq:AM_sol} is linear and convex with respect to $\boldsymbol{\gamma}$ for a given $f_\text{AM}$, it is non-convex with respect to $f_\text{AM}$ \cite{Bounds_4Paramfit_Harmonics}. Thus, for short window lengths (e.g., 60 ms) and noisy signals, the algorithm may converge on a combination of incorrect parameters with a smaller residual error than the ground truth. The GSS can be "guided" towards the global minimum by applying knowledge of the model: $a_\text{AM}\in [0,0.5]$ and $A_0>0$. Thus, at each GSS iteration, a feasibility check determines whethe the estimated parameters satisfy:
\begin{equation}
    \gamma_0>0 \text{ and }0\leq a_\text{AM}=\gamma_0^{-1}\sqrt{\gamma_1^2+\gamma_2^2}\leq 0.5.
\end{equation}
If the minimum of $\{\rho_\text{AM}(f_1), \rho_\text{AM}(f_2)\}$ corresponds to an infeasible parameter set, then the reverse direction is selected. If both are infeasible, the algorithm stops and the remaining feasible parameters are examined. With this rule, the estimated parameters are restricted to a feasible region. This specific configuration of GSS and LS will be denoted as $GSS/LS_\text{AM}$.

\subsubsection{Amplitude Step Detection}\label{sec:AS}
Abrupt changes in the amplitude (or phase) of the AC waveform are often due to faults, circuit breaker operations or network reconfigurations. The discontinuity in the waveform results in a scattering of energy across the frequency spectrum, yielding large errors in phasor estimations \cite{BeyondPhasors,RedefineFreq_PhaseSteps}. Instead of extracting an underlying fundamental tone, which has little meaning during such an event \cite{Roadmap_CPOW}, the dynamic is modeled as a transition from one state (either SS or dynamic) to another: once the step is detected, pre- and post-states are estimated separately.

This paper opted for a straightforward approach to step detection, amidst various algorithms proposed in the literature, such as Kalman filtering \cite{Kalman} and wavelet analysis \cite{step_detector_mario}. The strategy identifies significant variations in the envelope and argument, which could impact the AM and PM/FR analysis. An AS is detected by analyzing the differential amplitude, $\Delta x_A(t_l)=x_A(t_l)-x_A(t_{l-1})$, for consecutive samples of the extracted envelope. A moving average over $L_{\Delta A}$ samples of the differential magnitude is computed simply as $\Bar{\Delta x_A}=(x_A(t_l)-x_A(t_{l-L_{\Delta A}}))/L_{\Delta A}$. 
A step is flagged for the period $[t_{AS,lb},t_{AS,ub})$, when the differential magnitude first deviates from the mean by a significant margin at $t_{AS,lb}$ (i.e., $|\Delta  x_A (t_{AS,lb})-\Bar{\Delta x_A}|>\epsilon_{\Delta A}$) to when it returns to within the accepted range at $t_{AS,ub}$. The sample with the maximum deviation from the mean (i.e., the steepest slope) is set as the step location $t_\text{AS}$. The pre-step envelope is the reported AM model from before the step was detected, appropriately shifted by the reporting time. When sufficient samples $L_A$ after $t_{AS,ub}$ are available, the post-step amplitude $A_0^{post}$ is estimated by a running average on the envelope over $L_A$ samples, $\Bar{x}_A$. 
The step depth is found by comparing the new $A_0^{post}=\Bar{x}_A$ to the pre-step value, an AM with a central magnitude of $A_0^{pre}$: $a_\text{AS}=(A_0^{post}-A_0^{pre})/A_0^{pre}$. Refer to Section \ref{sec_design_parameters} for details on the threshold and moving average length selections.

\subsection{Argument Analysis}
The instantaneous argument of \eqref{eq:analyticform} is:
\begin{equation}%x_A(t)=|\Tilde{x}|=|x_A(t)e^{j\phi(t)}|
    x_\phi(t_l)=\angle \Tilde{x}(t_l)=2\pi f_0 t_l+g_\varphi (t_l)+\varphi_0.
\end{equation}
 The argument reflects disturbances like phase steps and embeds information on the frequency variation of the waveform. If AS and PS are not detected in the window (see \ref{sec:AS} and \ref{sec:PS}), the phase variation is assumed to be either a FR, PM, or SS. The SS case can be approximated as either of the former with a very low ramp rate or modulation depth, respectively.

\subsubsection{Frequency Ramps}\label{sec:FR}

The estimation of the argument parameters is again formulated as a LS problem ($LS_\text{FR}$) where the assumed model of the instantaneous phase is:
\begin{equation}\label{eq:FRmodel}
   x_\text{FR}(t_l)=\beta_0+\beta_1t_l+\beta_2 t_l^2.
\end{equation}
where $\beta_0=\varphi_0$, $\beta_1=2\pi f_0$, $\beta_2=R\pi$.
The objective function is therefore:
\begin{equation}
   \underset{\boldsymbol{\beta}}{\arg\min} \lVert \textbf{D}_\text{FR}\boldsymbol{\beta}-\textbf{x}_\phi\rVert_2
\end{equation}
where 
$\boldsymbol{\beta}=\begin{bmatrix}
\beta_0,\beta_1,\beta_2 \end{bmatrix}^T$, $
\mathbf{x}_\phi = \begin{bmatrix} x_\phi(t_0),\hdots x_\phi(t_{L-1})\end{bmatrix}^T$
and
\begin{equation}
\textbf{D}_\text{FR} =\begin{bmatrix}
    1 & t_0 & t_0^2 \\
    \vdots & \vdots & \vdots \\
     1 & t_{L-1} & t_{L-1}^2
    \end{bmatrix}.
\end{equation}
Note that the input vector, $\textbf{x}_\phi$, is the unwrapped argument of the analytic signal. The solution can then be found as:
\begin{equation}\label{eq:FR_sol}
    \boldsymbol{\beta}^*=\boldsymbol{\Lambda}_\text{FR}^{-1}\textbf{D}_\text{FR}^T \textbf{x}_\phi
\end{equation}
where, if the Gram matrix $\boldsymbol{\Lambda}_\text{FR}=(\textbf{D}_\text{FR}^T\textbf{D}_\text{FR})$  is invertible, the solution is unique. Indeed, for $L>0$ and $T_s>0$, $\boldsymbol{\Lambda}_\text{FR}$ is positive semi-definite and, therefore, its inverse always exists. 
For the reconstructed argument $x_\text{FR}(t_l)$, derived from \eqref{eq:FRmodel} and the solution to \eqref{eq:FR_sol}, the residual error is:
\begin{equation}
    \rho_\text{FR}(\boldsymbol{\beta})=\sum_{l=0}^{L-1}(x_\text{FR}(t_l,\boldsymbol{\beta})-x_\phi(t_l))^2.
\end{equation}   
\subsubsection{Phase Modulations}\label{sec:PM}
As with AM, modulations in the phase are often low-frequency and can be challenging to detect in short windows. The instantaneous argument is modeled as:
\begin{equation}
\begin{split}
    x_\text{PM}&(t_l)=\\
    &\nu_0+\nu_1 t+\nu_2 sin(2\pi f_\text{PM} t_l)+\nu_3 cos(2\pi f_\text{PM} t_l)
    \end{split}
\end{equation}
where $\varphi_0=\nu_0$, $2\pi f_0=\nu_1$, $a_\text{PM}=\sqrt{\nu_2^2+\nu_3^2}$ and $\varphi_\text{PM}=atan(\nu_3/\nu_2)$.
The objective function is:
\begin{equation} \label{eq:pm_OBJ}
   \underset{\boldsymbol{\nu}}{\arg\min} \lVert \textbf{D}_\text{PM}(f_\text{PM})\boldsymbol{\nu}-\textbf{x}_\phi\rVert_2
\end{equation}
where $\boldsymbol{\nu}^T=\begin{bmatrix}
\nu_0 & \nu_1 & \nu_2 & \nu_3
\end{bmatrix}$,
and 
\begin{equation}
\begin{split}
&\textbf{D}_\text{PM}(f_\text{PM}) =\\
&\begin{bmatrix}
    1 & t_0 & sin(2\pi f_\text{PM} t_0) &  cos(2\pi f_\text{PM} t_0)\\
    \vdots & \vdots & \vdots & \vdots \\
     1 & t_{L-1} & sin(2\pi f_\text{PM} t_{L-1}) & cos(2\pi f_\text{PM} t_{L-1}) 
    \end{bmatrix}
    \end{split}
\end{equation}

The 4-parameter solution for a fixed $f_\text{PM}$ is therefore:
\begin{equation} \label{eq:PM_sol}
    \boldsymbol{\nu}^*(f_\text{PM})=\boldsymbol{\Lambda}_\text{PM}^{-1}\textbf{D}_\text{PM}(f_\text{PM})^T\textbf{x}_\phi
\end{equation}
where the solution is unique if the Gram matrix $\boldsymbol{\Lambda}_\text{PM}=(\textbf{D}_\text{PM}(f_\text{PM})^T\textbf{D}_\text{PM}(f_\text{PM}))$ is invertible.  As with AM detection, a Golden Section Search is run to identify $f_\text{PM}$ in a specified range (e.g., $[1,5]$~Hz) where, at each iteration, the solution to \eqref{eq:PM_sol} and the corresponding residual are calculated:
\begin{equation}\label{eq:PM_res}
    \rho_\text{PM}(f_\text{PM})=\sum_{l=0}^{L-1}(x_\text{PM}(t_l,f_\text{PM},\boldsymbol{\nu}(f_\text{PM}))-x_\phi(t_l))^2
\end{equation}
Note that $x_\text{PM}(t_l)$ is the reconstructed waveform for a given $f_\text{PM}$ and $\boldsymbol{\nu}(f_\text{PM})$, computed via \eqref{eq:PM_sol}. This specific configuration of GSS and LS is denoted as $GSS/LS_\text{PM}$, but is similar to $GSS/LS_\text{AM}$. Since \eqref{eq:pm_OBJ} is nonlinear and nonconvex in $f_\text{PM}$, it is possible for the $GSS/LS_\text{PM}$ algorithm to converge to a local minimum \cite{Bounds_4Paramfit_Harmonics}. As in \ref{sec:AM}, a feasibility check is performed to ensure that $a_\text{PM}\leq \pi/2$, which is considered a reasonable range for modulations in power systems \cite{entsoe2017-EUoscillation}:%subsync_oscillations
\begin{equation}
    a_\text{PM}=\sqrt{\nu_2^2+\nu_3^2}\leq \pi/2.
\end{equation}

	\subsubsection{Phase Step Detection}\label{sec:PS}
A PS represents an abrupt change in the instantaneous argument of the waveform and, therefore, can be detected by analyzing the difference between consecutive samples of the argument: $\Delta x_\phi(t_l)=x_\phi(t_l)-x_\phi(t_{l-1})$. As in \ref{sec:AS}, the differential of the argument is fed into a $L_{\Delta\phi}$-sample running average filter. %The length $N_{\Delta\phi}=32$ was selected to be highly responsive to avoid triggering on phase modulations. 
When the differential deviates from the average $\Bar{\Delta x_\phi}$ by a significant margin 
(i.e., $|\Delta  x_\phi (t_{\text{PS},lb})-\Bar{\Delta x_\phi}|>\epsilon_{\Delta \phi}$ where $\epsilon_{\Delta\phi}$ is a threshold value), a phase step is flagged for the period $[t_{\text{PS},lb},t_{\text{PS},ub})$. The sample with the maximum deviation from the mean is set as the step location $t_\text{PS}$. Estimation of the phase step magnitude will be discussed in more detail in the following section.

\subsection{FBA Overview}\label{sec:FBAoverview}

Algorithm \ref{alg_FBA} outlines the full FBA method by connecting the components discussed in Sections \ref{sec:AM} through \ref{sec:PS}. In line 1, the envelope and argument are extracted from the analytic signal.
In line 2, a running mean on the differential envelope and phase is used to check if an AS or PS is present. If neither are detected (line 3), $GSS/LS_\text{AM}$ (\ref{sec:AM}) is applied to the envelope to estimate AM parameters in line 5, and reconstruct the envelope $x_\text{AM}(t_l)$ (line 6). 

Simultaneously, the $GSS/LS_\text{PM}$  and $LS_\text{FR}$ algorithms, from Sections \ref{sec:PM} and \ref{sec:FR}, are applied to estimate the parameters of the instantaneous argument (lines 8 and 9). The residual errors, $\rho_\text{PM}$ and $\rho_\text{FR}$, are computed and compared. It is notable that, when a FR is present, $\rho_\text{PM}\lessapprox \rho_\text{FR}$ may occur,
due to the additional degrees of freedom in the PM model (i.e., $a_\text{PM}$ and $f_\text{PM}$) and the masking effect of noise. Heuristically, it was decided that, if the residuals are within a tolerance (i.e., $|\rho_\text{PM}-\rho_\text{FR}|<\epsilon_\text{FR}$), FR is selected as the best approximation (lines 10 and 11). While the reverse is also possible (i.e., a PM falsely identified as a FR), it is less common. If the residual errors are sufficiently distinct, the dynamic with the smaller error is selected to represent the phase variation (lines 13-16).

If the mean differential of the envelope or argument exceeds the selected thresholds at any point in the window, a step transient is flagged and the point of maximum deviation is set as the amplitude or phase step time (lines 18 and 19). The time when both $\Delta x_A(t_l)$ and $\Delta x_\phi(t_l)$ return and stay within their respective thresholds is set as $t_{ub}$ (line 20).

\begin{algorithm}[]
\caption{Functional Basis Analysis Method}
\label{alg_FBA}
\begin{algorithmic} [1]
\REQUIRE $[x(t_l)]$
\STATE{$\Tilde{x}(t_l)=x(t_l)+\mathcal{H}(x(t_l)) \rightarrow x_A(t_l), x_\phi(t_l)$}
%\vspace{.3cm}
\STATE{$\Bar{\Delta x_A}=\mean[\Delta x_A(t_l)]_{L_{\Delta A}}$, $\Bar{\Delta x_\phi}=\mean[\Delta x_\phi(t_l)]_{L_{\Delta \phi}}$}

\IF[No step detected]{$|\Delta  x_A (t_l)-\Bar{\Delta x_A}|\leq\epsilon_{\Delta A} \wedge |\Delta  x_\phi (t_l)-\Bar{\Delta x_A}|\leq\epsilon_{\Delta\phi}$}
\STATE{Envelope Analysis:}
\STATE{$GSS/LS_\text{AM}\rightarrow f_\text{AM}^*,\boldsymbol{\gamma}^*\rightarrow x_\text{AM}(t_l)$} 
    \STATE{$\hat{x}_A^{[r]}(t_l)=x_\text{AM}(t_l)$}
   
    \STATE Argument Analysis:

    \STATE{$GSS/LS_\text{PM}\rightarrow f_\text{PM}^*,\boldsymbol{\nu}^*\rightarrow x_\text{PM}(t_l),\rho_\text{PM}$}
    
    \STATE{$LS_\text{FR}\rightarrow \boldsymbol{\beta}^*\rightarrow x_\text{FR}(t_l), \rho_\text{FR}$}
    \IF[PM $\approx$ FR]{$|\rho_\text{FR}-\rho_\text{PM}|<\epsilon_\text{FR}$}
        \STATE{$\hat{x}_\phi^{[r]}=x_\text{FR}(t_l)$} \COMMENT{Select FR}
    \ELSE
        \IF[Compare PM and FR]{$\rho_\text{FR}<\rho_\text{PM}$}
            \STATE{$\hat{x}_\phi^{[r]}(t_l)=x_\text{FR}(t_l)$}
        \ELSE
            \STATE{$\hat{x}_\phi^{[r]}(t_l)=x_\text{PM}(t_l)$}

        \ENDIF
   \ENDIF

\ELSE[Step detected]
    \STATE{$t_\text{AS}=\argmax_{t_l}(|\Delta  x_A(t_l)-\Bar{\Delta x_A}|)$}
    \STATE{$t_\text{PS}=\argmax_{t_l}(|\Delta  x_\phi(t_l)-\Bar{\Delta x_\phi}|)$}
    
    \STATE{$t_{ub}= \inf \{ t : \forall t_l > t,$ \text{s.t. ...}\\  $\vert\Delta  x_A (t_l)-\Bar{\Delta x_A}|\leq\epsilon_{\Delta A} \wedge |\Delta  x_\phi (t_l)-\Bar{\Delta x_\phi}|\leq\epsilon_{\Delta\phi}\}$}
   
     \STATE Envelope Analysis:

    \IF[Estimate post-step $A_0$]{$L-t_{ub}/T_s\geq L_A$} %post  step estimate availablee
        \STATE{$A_0^{post}=\mean[x_A(t_l>t_{ub})]_{L_{A}}$}
        \STATE{$\hat{x}_A^{[r]}(t_l)\leftarrow$\eqref{AS_piecewise}}
   \ELSE[Use past estimate]
        \STATE{$\hat{x}_A^{[r]}(t_l)\leftarrow$ \eqref{AM_shifted}}
        
    \ENDIF
    \STATE Argument Analysis:
    \IF{$t_{ub}$ is not null} %post  step estimate availablee
        \IF[Estimate PS depth]{$L-t_{ub}/T_s< L_\text{FR}$}
    %post  step estimate availablee
            \STATE{$a_\text{PS}=x_\phi(t_{ub})-\hat{x}_\phi^{[r-1]}(t_{ub}+T_{RR})$}
            \STATE{$\hat{x}_\phi^{[r]}(t_l)\leftarrow$\eqref{PS_piecewise}}
        \ELSE[Estimate post-step FR]  %post-step state (FR) is available
            \STATE{$LS_\text{FR}(x_\phi(t_l\geq t_{ub}))\rightarrow \boldsymbol{\beta}^{*post}\rightarrow x_\text{FR}^{post}(t_l-t_{ub})$}
\STATE{$\hat{x}_\phi^{[r]}(t_l)\leftarrow$\eqref{PSFR_piecewise}}
            
        \ENDIF

    \ELSE[Use past estimate]
        \STATE{$\hat{x}_\phi^{[r]}(t_l)\leftarrow$\eqref{FR_shifted}}
        
    \ENDIF
\ENDIF

\STATE {$\hat{x}^{[r]}(t_l)=\hat{x}_A^{[r]}(t_l)cos(\hat{x}_\phi^{[r]}(t_l))$}\COMMENT{Reported model}

\end{algorithmic}
\end{algorithm}
Estimating the post-step amplitude, frequency variation and PS depth, depends on the number of available post-step samples. For example, if fewer than $L_A$ samples are present after $t_{ub}$ (line 25-26), the estimated envelope is the AM estimate $\hat{x}_A^{[r-1]}(t_l)$ from the previous report, $r-1$, before the step. The predicted parameters for report $r$ need to be shifted by the reporting time $T_{RR}=1/F_{RR}$. If the previous reported envelope was $\hat{x}_A^{[r-1]}(t_l)$, the predicted model is:
\begin{equation}\label{AM_shifted}
    \hat{x}_A^{[r]}(t_l)=\hat{x}_A^{[r-1]}(t_l+T_{RR}).
\end{equation}
 Once $\Delta x_A(t_l)$ and $\Delta x_\phi(t_l)$ return to within threshold and the new mean amplitude is available, the post-step state is updated based on the magnitude and location of the step (lines 22-24). The estimated model is the piecewise function: 
\begin{equation}\label{AS_piecewise}
\begin{split}
    \hat{x}_A^{[r]}&(t_l)=\\
    &\hat{x}_A^{[r-1]}(t_l+T_{RR}) h(t_\text{AS}-t_l) +A_0^{post}h(t_l-t_\text{AS}).
    \end{split}
\end{equation}

As the step moves through the window, the pre-step model is shifted appropriately, while the post-step envelope estimate is continuously updated as more samples are acquired. Similarly, when an AS/PS is detected, the previous argument estimate is assumed to be unaffected by the step. If $\Delta x_A(t_l)$ and $\Delta x_\phi(t_l)$ remain outside of the threshold (lines 35 and 36), the previous estimate is appropriately shifted to predict the current state: 
\begin{equation}\label{FR_shifted}
\hat{x}_\phi^{[r]}(t_l)=\hat{x}_\phi^{[r-1]}(t_l+T_{RR})    
\end{equation}
If both $\Delta x_A$ and $\Delta x_\phi$ return to and stay within their respective thresholds, at $t_{ub}$, the $LS_\text{FR}$ estimator begins accumulating samples. At this time, the PS depth is also approximated by comparing the predicted argument at $t_{ub}$ to the instantaneous argument of the analytic signal (lines 29-31):
\begin{equation}
   a_\text{PS}=x_\phi(t_{ub})-\hat{x}_\phi^{[r-1]}(t_{ub}+T_{RR}).
\end{equation}
The argument model is adjusted accordingly in line 31:
\begin{equation}\label{PS_piecewise}
\hat{x}_\phi^{[r]}(t_l)=\hat{x}_\phi^{[r-1]}(t_l+T_{RR})+a_\text{PS} h(t_l-t_\text{PS}).
\end{equation}
Finally, once there are enough post-step samples (i.e., at least $L_\text{FR}$ samples following $t_{ub}$), the $LS_\text{FR}$ estimator approximates the new frequency dynamic, $x_\text{FR}^{post}(t_l-t_{ub})$, with the time axis starting at $t_{ub}$. The reconstructed waveform in line 34 is, therefore, a piece-wise function based on the shifted pre-step estimate, the step location $t_\text{PS}$, and the new post-step model (also appropriately shifted to start at the step location): 
\begin{equation}\label{PSFR_piecewise}
\begin{split}
    \hat{x}_\phi^{[r]}(t_l)=&\hat{x}_\phi^{[r-1]}(t_l+T_{RR}) h(t_\text{PS}-t_l) +\hdots \\
    &x_\text{FR}^{post}(t_l-(t_{ub}-t_\text{PS})) h(t_l-t_\text{PS}).
    \end{split}
\end{equation}

At last, in line 37, the complete signal model is reported.

\subsection{Selecting Design Parameters}\label{sec_design_parameters}

While the FBA algorithm's design parameters are adaptable to specific applications, justification of the specific values selected for this implementation are provided below.
\subsubsection{Step detection}
For this implementation, the step detection thresholds are set at $\epsilon_{\Delta A}=0.035 \%$ (with respect to the analog input voltage range) %0.0035V 
and $\epsilon_{\Delta A}=0.0002\pi$. The length of both differential mean filters is selected to be $L_{\Delta A}=L_{\Delta \phi}=32$. This combination of threshold and moving average length is chosen to avoid false triggering on signals with a noise level of 60 dB and during the most extreme AM and PM dynamics tested (i.e., depth of 0.5 and 5 Hz frequency). The moving average on the envelope has a length $L_A=64$ samples for improved  accuracy though a slower response time relative to the other moving averages. Finally, the moving average lengths are chosen as a power of 2 for easy bit-wise manipulation in fixed-point calculations. 

\subsubsection{GSS settings} \label{GSS_param}
The initial interval $[f_{lb},f_{ub}]=[1,5]$ Hz of the GSS algorithm is selected to cover the range of AM and PM expected in power systems, as inspired by the PMU Standard \cite{PMUStandard}. Frequencies below this range were found to be too slow to accurately characterize with a 60 ms window and will not significantly impact the reported error values.

Furthermore, it was determined that five total iterations of the GSS are sufficient for parameter convergence. For justification, residuals at the evaluated frequencies, $f_\text{AM}$ or $f_\text{PM}$, were compared using 60 ms windowed signals with 60 dB Gaussian noise. After 5 iterations, the differences in the residuals were on the order of $10^{-6}$, which is unlikely to be distinguishable in a fixed-point implementation. Even for high-precision calculations, additional iterations served to minimize noise in the signal rather than improve the estimation.

\subsubsection{FR accuracy}
The uncertainty in the FR estimation of $\beta_0$, $\beta_1$ and $\beta_2$ is inversely proportional to $L$, $L^3$ and $L^5$, respectively (see the Appendix for the Cramer-Rao Lower Bound). For shorter window lengths and when analyzing noisy signals, the parameter error increases. Therefore, a minimum sample number $L_\text{FR}$ is selected, below which the FR estimates are considered unreliable.
The minimum number of samples required for valid FR estimations was determined to be $L_\text{FR}=300$ samples for $F_s=10$ kHz. This was decided by evaluating the maximum frequency error of the estimator over 1000 random noise seeds at 60 dB for $L=1...600 $ samples. The FE for $L\geq300$ samples (i.e., 30 ms) satisfies the PMU Standard limit of 5 mHz for SS signals \cite{PMUStandard}.

%% file: Chapters/3_LVImplementation.tex
\section{Embedded Hardware Implementation} \label{sec:LVimplementation}

As in other hardware implementations of PMUs \cite{Derviskadic2020_iIpDFT_FPGA}, the FBA algorithm is deployed in an FPGA-based (Field Programmable Gate Array) platform, which can perform high speed logic and computations with deterministic timing, making it ideal for real-time signal processing. This section presents the adaptations of the FBA to ensure efficient FPGA resource allocation and minimum latency, including the design and architecture of the Hilbert filter, derivation of the recursive estimation, and storage of the AM/PM bases.

\subsection{Approximating the Analytic Signal}\label{sec:HilbertFilter}

The practical implementation of the Hilbert transform requires approximations of the ideal frequency response, \eqref{eqn_HT_freq}, as the impulse response is non-causal. For this work, finite impulse response (FIR) filters are selected to approximate the complex analytic signal since they have a linear phase response, constant group delay and low sensitivity to coefficient quantization \cite{HilbertFilter_guide}. While the large delay and the computational complexity of FIRs is a drawback, the quality of the output analytic signal is ideal for broadband analysis. 

The frequency response of an ideal Hilbert filter is:
\begin{equation}
    H(e^{j\omega})=\left\{
\begin{array}{ll}
      -j & 0\leq\omega\leq \pi \\
      j & -\pi\leq \omega\leq 0 \\
\end{array} 
\right. 
\end{equation}
with unity magnitude and $\mp \pi/2$ phase shift for $\pm$ frequencies. 

FIR Hilbert filters are antisymmetric and are classified as either type III (odd length) or type IV (even length) filters. With the  Parks-McClellan technique \cite{parksmcclellan}, the optimum equiripple Hilbert filter can be designed to specification:
\begin{equation}
    (1-\delta)\leq |H(e^{j\omega})|\leq (1-\delta) \text{ for } \omega_{lb}\leq\omega\leq\pi-\omega_{lb}
\end{equation}
where $\omega_{lb}$ is the lower pass-band frequency and $\delta$ is the pass-band ripple. However, analysis of power grid waveforms operating near nominal frequencies (e.g., 50 or 60 Hz) would require an FIR Hilbert filter with a very high filter length $L_H$ and a prohibitive number of coefficients in order to achieve
the necessary transition width $\omega_{lb}$ and low ripple \cite{PM_length}.

\begin{figure}[!t]
\centering

\includegraphics[width=\linewidth]{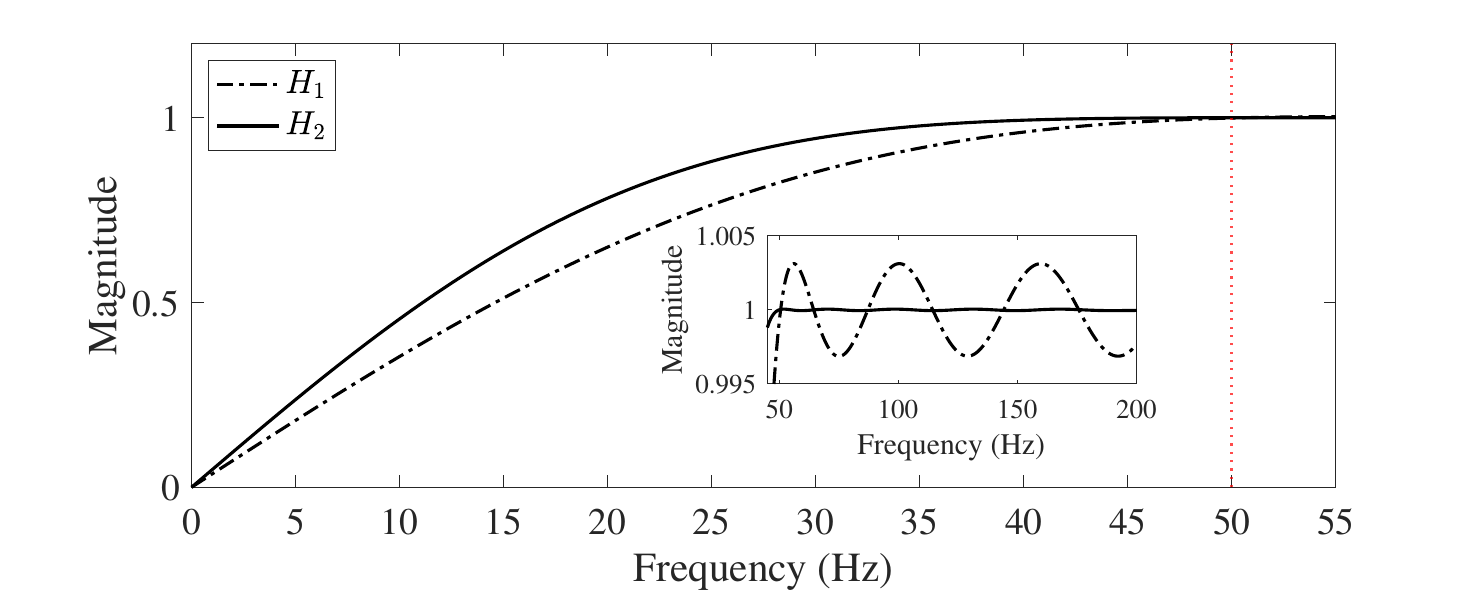}
\caption{Magnitude response for filters $H_1$ and $H_2$ with $F_s=10$ kHz and $\omega_{lb}=50$ Hz (red). \label{fig_bode}}
\end{figure}

Fortunately, several alternatives to the Parks-McClellan technique exist to build efficient and sharp Hilbert filters. This paper uses the Frequency Transformation (FT) \cite{FT} method\footnote{Special thanks to Professor David Romero at the Autonomous University of Quintana Roo State, Mexico, for his help in developing the deployed filters.}, which is based on the cascaded interconnection of identical subfilters. The subfilter, $G(z)$, a type III Hilbert filter, is designed to have a relaxed ripple requirement but strict transition width. The number of subfilter copies and the coefficients connecting the duplicates depend on the design of a second, type IV Hilbert filter, the prototype filter $P(z)$. This filter, in contrast, has a strict ripple requirement and enlarged transition width. By relaxing one or the other specification, both filters have low complexity and reduced order. The combined filter $H(z)$ preserves the magnitude response of the prototype filter, while the frequency domain is modified by the subfilter \cite{FT}.

For this paper, two combined filters, shown in Table \ref{tab:Hilbert}, were developed to compare performance and latency. For each, the lengths of their subfilter and prototype filters, $L_G$ and $L_P$, respectively, were found using the steps presented in \cite{FT} and \cite{Est_Min_Num_Dist_Multipliers_FTFILT} to satisfy the performance specifications. Filter $H_1$ is a low order filter with a relaxed ripple constraint while Filter $H_2$ is a high performance, higher latency filter. The frequency response for both filters is compared in Fig. \ref{fig_bode}.

\begin{table} []
    \caption{Design of Hilbert filters based on the FT method.}
    \centering
    \begin{tabular}{ccccccc} \midrule 
         &    $\omega_{lb}$&$\delta$&$L_P$&  $L_G$& Multipliers &Delay (Samples)\\ \midrule 
         $H_1$&    $0.01\pi$&0.004& 16 &  27 &  15 & 210 \\ \midrule 
         $H_2$&    $0.01\pi$&0.0001&22&  37& 20 &399\\ \midrule
    \end{tabular}
    \label{tab:Hilbert}
\end{table}

The FT-Hilbert filters designed above can be implemented using the Pipelining-Interleaving (PI) technique \cite{PI_filters, FT}, an area-efficient, multirate filter architecture which multiplexes a repeated filter. Specifically, the configuration in \cite{PI_filters} is applied such that a single subfilter is run at a higher rate, maintaining high throughput while decreasing FPGA consumption. The total group delay of this PI-FT-Hilbert filter, in samples, is:
\begin{equation}
\text{Delay}=(L_G-1)/2+1+(L_G+1)(L_P/2-1),
\end{equation}
For $F_s=10$ kHz, the delay of $H_2$ is $T_{H_2}=39.9$ ms, well within the limit of 140 ms for M-class PMUs \cite{PMUStandard}. 

\subsection{Recursive Frequency Ramp Estimation}
Modeling the instantaneous phase as a FR (see Section \ref{sec:FR}), involves solving \eqref{eq:FR_sol}. However, calculating the matrix inverse of $\boldsymbol{\Lambda}_{FR}(L,T_s)$ is computationally expensive and sensitive to fixed-point arithmetic accuracy. Instead, the 6 unique coefficients of the symmetric $\boldsymbol{\Lambda}_{FR}^{-1}$ are computed (for select $L$ and $T_s$) in double precision then converted to a fixed-point representation and stored for easy access and high precision. Next, the component $\textbf{D}_{FR}^T\textbf{x}_\phi$ can be expressed as:

\begin{equation}
\begin{split}
    \textbf{D}_{FR}^T\textbf{x}_\phi=
    \begin{bmatrix}
        \sum_{l=0}^{L-1} x_\phi(t_l)\\
        \sum_{l=0}^{L-1} t_lx_\phi(t_l)\\
        \sum_{l=0}^{L-1} t_l^2x_\phi(t_l)\\
    \end{bmatrix} & =\begin{bmatrix}
       s_0\\
        T_s s_1\\
        T_s^2 s_2\\
    \end{bmatrix} 
    \end{split}
\end{equation}
where $s_\lambda=\sum_{l=0}^{L-1}i^
\lambda x_\phi(t_l)$. A recursive formulation of this term enables streaming updates and reduced complexity. Let $s_\lambda(t_n)=\sum_{l=0}^{L-1}l^\lambda x_\phi(t_{(n-L+1)+l})$, 
%$s_\lambda(t_n)=\sum_{l=0}^{L-1}l^\lambda x_\phi(t_{q+l})$, where $q=n-L+1$ 
for a window size $L$ with $x_\phi(t_n)$ being the most recent sample. Therefore,
\begin{equation} \label{eq:S0}
    s_0(t_n)=s_0(t_{n-1})+x_\phi(t_{n}) -x_\phi(t_{n-L}),
\end{equation}
\begin{equation}\label{eq:S1}
\begin{split}
    s_1(t_n)=&s_1(t_{n-1})-s_0(t_{n-1})+\hdots \\
    &x_\phi(t_{n})(L-1) + x_\phi(t_{n-L}),
\end{split}
\end{equation}
\begin{equation}\label{eq:S2}
\begin{split}
    s_2(t_{n})=&s_2(t_{n-1})-2s_1(t_{n-1})+s_0(t_{n-1})+\hdots \\
    &x_\phi(t_{n})(L-1)^2 - x_\phi(t_{n-L}).
\end{split}
\end{equation}
While the direct approach would  require $2(L-1)$ additions and $2L$ multiplications to calculate $\textbf{D}_{FR}^T\textbf{x}_\phi$, the recursive formulation is much simpler. The complexity of each operation is summarized in Table \ref{tab_complexity} for $F_s=10$ and $L=600$ samples or 60 ms, to comply with the P-class PMU Standard. 

\begin{figure}[]
\centering
\includegraphics[width=0.9\linewidth]{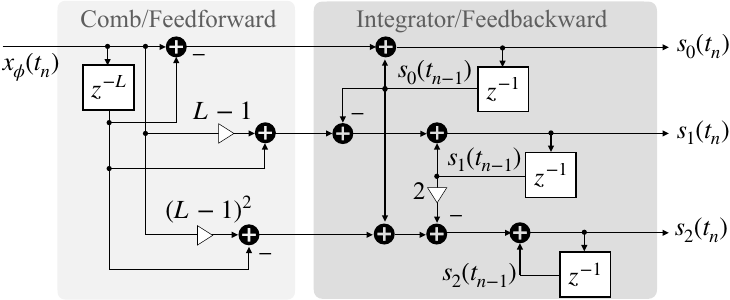}
\caption{Block diagram for recursive moving accumulators $s_\lambda$.}
\label{fig_CIC}
\end{figure}

The calculation of $\textbf{D}_{FR}^T\textbf{x}_\phi$ becomes a recursive, sliding accumulator with distinct similarities to Cascaded Integrator-Comb filters (CIC) \cite{CIC_filters}, involving a feed-forward and feedback portion, as shown in Fig. \ref{fig_CIC}. CIC filters are guaranteed stable when implemented in fixed-point, rather than floating point, as the arithmetic operations are performed exactly \cite{CIC_filters}.

For parameter estimation, the argument $x_\phi(t_l)$ should be unwrapped but bounded for fixed-point implementation. The argument is automatically wrapped between $-\pi$ and $\pi$ and saved to a circular buffer. A phase wrap is detected by comparing each argument value to the previous value (e.g., when $x_\phi(t_{n})=1.01\pi$ is wrapped to $-0.99\pi$, $x_\phi(t_{n})-x_\phi(t_{n-1})\leq-\pi$). When a wrap is detected: 1) the index of the circular buffer address is saved, 2) the wrap count is incremented $W+1$, and 3) the current sample is unwrapped $x_\phi(t_{n})=x_\phi(t_{n})+2\pi W$. When the beginning of the window reaches a wrap index (i.e., a wrap is "cleared"): 1) the recorded index of the circular buffer is forgotten, 2) the wrap count is decremented $W-1$, and 3) the accumulating sums, $s_\lambda$, are adjusted by a correction term in order to "shift" the full window by $-2\pi$. Specifically  $s_\lambda=s_\lambda - M_\lambda$ where $M_0=2\pi(L-1)$, $M_1=\pi(L-2)(L-1)$ and $M_2=\pi(L-2)(L-1)(2L-3)/3$.

\subsection{Amplitude/Phase Modulation Bases}

The GSS algorithm, applied for both AM and PM estimation, iteratively approaches the modulating frequency yielding the minimum residual. At each iteration, the error is evaluated at four frequencies $\{ f_{lb}, f_1, f_2, f_{ub}\}$, which are subsequently adjusted using the golden ratio for maximal efficiency \cite{GSS}. Section \ref{sec_design_parameters} justifies why the five total GSS iterations are sufficient for parameter convergence. Consequently, exactly $2^5$ frequencies are possible in the evaluation. For the frequency range of $[1,5]$ Hz (see Section \ref{GSS_param}), only 26 frequencies exhibit distinct characteristics, as the remaining 6 kernels are nearly identical to neighboring kernels, differing by less than 0.07 mHz. Each AM and PM basis $\textbf{D}(f)$ is composed of:
\begin{itemize}
    \item an $L\times1$ vector $cos(2\pi f_{AM} t_l)$ or $cos(2\pi f_{PM} t_l)$
    \item an $L\times1$ vector $sin(2\pi f_{AM} t_l)$ or $sin(2\pi f_{PM} t_l)$
    \item a $6\times1$ or $10\times1$ vector containing the unique coefficients of the symmetric inverse $\boldsymbol{\Lambda}^{-1}(f_{AM})$ or $\boldsymbol{\Lambda}^{-1}(f_{PM})$.
\end{itemize}

\subsection{Device Specific Consumption and Latency}
The FBA is implemented in a National Instruments compactRIO (cRIO) 9039 \cite{NI_9039_manual}, an embedded controller with a Real-Time Processor and a reconfigurable Xilinx Kintex-7 FPGA. The NI 9215 Analog Input module \cite{NI_9215_manual} is used for signal acquisition, with a maximum input range of $\pm 10$ V, while synchronization to UTC time is achieved with the NI 9467 module \cite{NI_9467_manual}, a GPS antenna and a GPS splitter. Note that GPS is expected to have a time uncertainty $3\sigma$ of $\pm 100$ ns when paired with commercial GPS receivers.

Table \ref{tab_complexity} lists the FPGA timing for each operation, with \textit{ticks} representing the FPGA time resolution of the 40 MHz onboard clock. For AM analysis, each 3-parameter estimation takes 1227 ticks (0.03 ms). To minimize resource use, the GSS iterations are performed sequentially for a total time of 0.24 ms. PM estimation is performed in parallel to AM analysis. Note that, the AM and PM estimations are only performed at the reporting rate $F_{RR}$ whereas the Hilbert filter and FR updates occur at the sampling frequency $F_s$.

The total computation time is $T_x=0.27$ ms. Therefore, the latency defined by the PMU Standard \cite{PMUStandard}, for the $H_2$ filter, $T_w=60$ ms, and $F_s=10$ kHz, is $T_x+T_H+T_w/2=70.13$~ms, well within the M class limit of 140 ms.

In terms of resource consumption, the Hilbert filter ($H_2$), and AM and PM bases use 0.7\%, 4.5\% and 4.5\%, respectively, of the available block memory (BRAM). In total, the FBA requires 9.4\% of the BRAMs (or 1,584 kbits), 23.4\% of the slice registers, 39.6\% of the DSP48s (e.g., multipliers) and 55\% of all available slices available on the cRIO 9039 FPGA. Therefore, ample space is available for channel expansion or the incorporation of additional applications.

\begin{table}[]
\centering
\caption{Computational Complexity and Latency. \label{tab_complexity}}
\begin{threeparttable}
\begin{tabular}{ccrrcc}
\toprule
\multirow{2}{*}{\textbf{}} & \multirow{2}{*}{\textbf{Operation}}  & \multicolumn{2}{c}{\textbf{FPGA Timing}} & \multicolumn{2}{c}{\textbf{Complexity}} \\%\multicolumn{3}{|c|}{Multicolumn Header}
\cmidrule(lr){3-4} \cmidrule(lr){5-6}
  &  & {Ticks} & {Time ($\mu s$)} & {$+/-$} & $\times$ \\

\midrule
\multirow{4}{*}{AM$^{\diamond}$} & $\textbf{D}_{AM}^T \textbf{x}_A^{\star\star}$  & $604$ & $15.10$ & $2(L-1)$& $2L$\\ 
                        & \eqref{eq:AM_sol}$^{\star\star}$ & $15$ & $0.38$ & $6$ & $9$ \\
                        & Residual$^{\star\star}$ & $607$ & $15.18$  & $2L$ & $4L$ \\ 
                        &  Feasibility$^{\star\star}$ & 1  & $0.03$  & $1$ & $3$                                  \\ \midrule
\multirow{4}{*}{PM$^{\diamond}$} & $\textbf{D}_{PM}^T \textbf{x}_\phi^{\star\star}$   & $605$  & $15.13$  & $3(L-1) $& $3L$\\
                        & \eqref{eq:PM_sol}$^{\star\star}$ & $17$  & $0.43$ & $16$  & $12$ \\  
                        &  Residual$^{\star\star}$ & $607$ & $15.18$ & $4L$  & $5L$ \\
                        &  Feasibility$^{\star\star}$  & 1 & $0.03$ & $1$  & $3$  \\ \midrule
\multirow{3}{*}{FR}  & $\textbf{D}_{FR}^T\textbf{x}_\phi^\star$  & $1$    & $0.03$   & $12$    & $4$     \\ 
& \eqref{eq:FR_sol}$^\star$ & 6 & 0.15 & 6& 9 \\
& Residual$^{\star\star}$ & 607 & 15.18 & $4L$& $3L$\\
\midrule
      & Filter $H_2^{\star}$                        & $847$                 & $21.18$                     & $38$                           & $20$                                 \\ 
\bottomrule
\end{tabular}
\begin{tablenotes}
\item $^\star$ Evaluated at assessment rate $F_{s}$. 
\item $^{\star\star}$ Evaluated at assessment rate $F_{RR}$. 
\item $^{\diamond}$ Per 3- or 4-parameter estimation.
\end{tablenotes}
\end{threeparttable}
\end{table}

%% file: Chapters/4_Experiments.tex
\section{Experimental Validation}\label{sec:Experimental_Validation}

This section assesses the performance of the FBA prototype device in controlled scenarios, using a PMU Calibration test-bed. Several static and dynamic tests from the PMU Standard \cite{PMUStandard} are replicated, as well as advanced multidynamic tests. The FBA is compared to static and dynamic phasor techniques, specifically the 3-point iterative Interpolated-DFT (i-IpDFT) \cite{iIPDFT} and Compressed Sensing Taylor-Fourier multifrequency (CSTFM) \cite{CompressiveSensing_TF_Frigo} method, respectively. 

\subsection{Experimental Setup}\label{sec_Calibrator_setup}

\begin{figure}[]
\centering
\includegraphics[width=\linewidth]{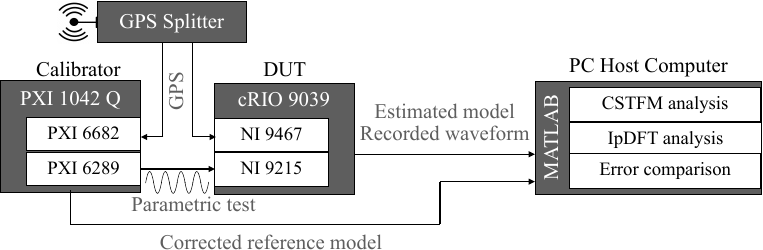}
\caption{Lab setup for validation tests.\label{fig_testsetup}
}
\end{figure}

 To evaluate the performance of a device under test (DUT), a well defined reference system is needed. As in \cite{AsjaThesis}, a PMU calibration system is used to generate test waveforms with precisely known reference magnitude, frequency, phase and Rate of Change of Frequency (ROCOF). 
 To evaluate state-of-the-art PMUs that are designed with a maximum Total Vector Error (TVE) of 0.01\% and 0.1\% in SS and dynamic scenarios, respectively, the calibrator must be able to report TVEs on the order of 0.00X\% and 0.0X\%, respectively\cite{Uncertainty_Calibration} \cite{powertechCalibrator_Danielle}. 

 The calibrator is implemented in a PXI industrial controller, capable of generating highly accurate waveforms thanks to its multi-device synchronization and dedicated triggering architecture. The calibrator consists of (1) a PXI-1042Q chassis \cite{NI_PXI1042Q_manual}, (2) a PXI-8110 Controller, (3) a PXI-6682 synchronization board, and (4) a PXI-6289 DAQ board with a $\pm 11$ V output limit. The PXI-6682 synchronizes the system to UTC time, via input from a GPS antenna, routed through a GPS splitter. Based on this reference clock, internal triggers are created corresponding to the standard pulse-per-second (PPS) and the user-specified digital-to-analog (DAC) sampling frequency of 500 kHz. The selected parameterized signal model is constructed and output from the PXI-6289 DAQ board at the start of the second. As shown in Figure \ref{fig_testsetup}, the DUT acquires the signal and timestamps the parameter estimates, allowing for alignment with the reference parameters.

Critically, a metrological characterization of the systematic error in the acquired signal must be conducted to correct the reference parameters. Both the generation and acquisition stages introduce non-trivial offsets in the magnitude and phase of the signals, due to temperature variation and the analog front ends of the ADC/DAC \cite{Uncertainty_Calibration,CharCalDelay}. To estimate these errors, 5 s of a 50 Hz steady-state sinusoid is acquired by the DUT. The waveform is divided into overlapping windows of 60 ms with 50 fps reporting rate, and the amplitude and phase ($A_{est}$, $\varphi_{est}$) are estimated by a phasor estimation algorithm known to have high accuracy in steady-state conditions\cite{iIPDFT}. The mean estimated amplitude and phase are used to compute a magnitude attenuation factor of $\zeta_{corr}=\mean(A_{est})/{A_0}=0.99987$ (i.e.,  $\approx 0.01\%$ attenuation), and a time delay of $T_{corr}=mean(\varphi_{est}-\varphi_0)/2\pi f_0=-3.98$ $\mu$s. 
To verify the validity of these corrections, the measured waveforms for each test are compared to the corrected reference model, yielding normal distributions of the time-domain residual errors that can be attributed to measurement noise  \cite{Uncertainty_Calibration}.

%% file: Chapters/5_Results.tex
\subsection{IEC/IEEE 60255-118 Compliance Tests}%IEC/IEEE 60255-118-1
\subsubsection{Single dynamic tests}\label{sec_single_dyn_test}
The compared methods are applied to several signal dynamic test waveforms mandated by the PMU Standard \cite{PMUStandard} as well as waveforms with parameters beyond the standard ranges to explore performance in extreme cases (e.g., amplitude modulations at depths of $a_{AM}=0.5$, frequency ramps at 5 and 10 Hz/s).
The tests are as follows (where  $A_0$ is relative to the voltage range of the DUT):
\begin{itemize}
    \item \textbf{SS}: $A_0=0.9$ p.u., $f_0=45:1:55$ Hz.
    \item \textbf{AM}: $A_0=0.6$ ~p.u., $a_{AM}\in\{10,50\}$\%, $f_0=50$ Hz, $f_{AM}=0.1:0.5:5$ Hz.
    \item \textbf{PM}: $A_0=0.9$ p.u., $ a_{PM}\in\{0.1,0.5\}$ rad, $f_0=50$ Hz, $f_{PM}=0.1:0.5:5$ Hz.
    \item \textbf{FR}: $A_0=0.9$ p.u., $f\in[45,55]$ Hz, $R=\{1,5,10\}$  Hz/s.
    \item \textbf{AS}: $A_0=0.6$ p.u., $f_0=50$ Hz,  $a_{AS}=10$ \%,
     \item \textbf{PS}: $A_0=0.6$ p.u., $f_0=50$ Hz, $a_{PS}=\pi/18$.
\end{itemize}
With the known ground truth parameters (see \ref{sec_Calibrator_setup}), error calculations can be made for all applied algorithms. To make a fair comparison with the phasor-based methods, the metrics specified by the PMU Standard \cite{PMUStandard} are reported: TVE, frequency error (FE) and ROCOF error (RFE), computed as the differential between consecutive frequency estimates. Additional metrics will be defined in the following sections for evaluation of performance during steps and multidynamics. 

The results, including performance for both Hilbert filters ($H_1$ and $H_2$), are presented in Table \ref{tab_IEEE}. Observe that, in most cases, FBA-$H_1$ yields higher errors due to the relaxed ripple specification of the filter. In SS tests, when $f_0<50$ the magnitude of the filter-generated imaginary component is attenuated, resulting in oscillations in the envelope and phase of the analytic signal which bias the FBA estimate. The stricter ripple requirements of $H_2$ mitigates this effect but does not entirely eliminated it, as is seen in the higher reported TVE. However, for $f_0\geq 48$ Hz, $H_2$ yields TVEs on the order of 0.00X \%. Possible solutions include correcting for the gain of the filter based on the frequency estimation, as done in \cite{cesar_correction}, or by designing a Hilbert filter with a smaller transition width.

\begin{table*}[]
\caption{Signal Dynamic Test Performance \cite{PMUStandard}.\label{tab_IEEE}}
\renewcommand{\arraystretch}{1}
\centering
\resizebox{0.9\linewidth}{!}{\begin{tabular}{cccccccc}
\toprule
\multirow{2}{*}{\textbf{Dynamic Test}} & \multirow{2}{*}{\textbf{Metric}} & \multicolumn{2}{c}{\textbf{Standard Limits}} & \multirow{2}{*}{\textbf{i-IpDFT}} & \multirow{2}{*}{\textbf{CSTFM}} & \multirow{2}{*}{\textbf{FBA ($H_1$)}} & \multirow{2}{*}{\textbf{FBA ($H_2$)}} \\ \cmidrule{3-4}
 &  & \multicolumn{1}{c}{{P class}} & {M class} &  &  &  &  \\ \midrule
 
\multirow{3}{*}{\textbf{SS}} & FE (mHz) & \multicolumn{1}{c}{5} & 5 & 0.1 & 0.3 & 6.8 & 0.2 \\  
 & RFE (Hz/s) & \multicolumn{1}{c}{0.4} & 0.1 & 0.012 & 0.008 & 0.260 & 0.010 \\
 & TVE (\%) & \multicolumn{1}{c}{1} & 1 & 0.005 & 0.007 & 0.700 & 0.060 \\ \midrule
 
\multirow{3}{*}{\textbf{\begin{tabular}[c]{@{}c@{}}AM \\ $a_{AM}=0.1/0.5 \text{ }p.u.$\end{tabular}}} & FE (mHz) & \multicolumn{1}{c}{60/-} & 300/- & 0.5/6.4 & 0.7/7.1 & 2.2/14.1 & 0.3/1.2 \\
 & RFE (Hz/s) & \multicolumn{1}{c}{2.3/-} & 14/- & 0.017/0.160 & 0.016/0.150 & 0.093/0.496 & 0.013/0.040 \\ 
 & TVE (\%) & \multicolumn{1}{c}{3/-} & 3/- & 0.600/5.200 & 0.050/0.400 & 0.070/0.310 & 0.020/0.170 \\ \midrule
 
\multirow{3}{*}{\textbf{\begin{tabular}[c]{@{}c@{}}PM \\ $a_{PM}=0.1/0.5\text{ }rad$\end{tabular}}} & FE (mHz) & \multicolumn{1}{c}{60/-} & 300/- & 17.0/86.0 & 41.0/220.0 & 3.0/7.5 & 1.3/4.1\\ 
 & RFE (Hz/s) & \multicolumn{1}{c}{2.3/-} & 14/- & 0.540/2.850 & 1.240/6.141 & 0.350/0.480 & 0.191/0.270 \\ 
 & TVE (\%) & \multicolumn{1}{c}{3/-} & 3/- & 0.540/2.700 & 0.040/0.280 & 0.080/0.280 & 0.010/0.020 \\ \midrule
 
\multirow{3}{*}{\textbf{\begin{tabular}[c]{@{}c@{}}FR   \\ $R=1/5/10\text{ }Hz/s$\end{tabular}}} & FE (mHz) & \multicolumn{1}{c}{10/-/-} & 10/-/- & 0.2/0.7/1.3 & 0.2/0.6/1.1 & 6.9/7.0/6.9 & 0.2/0.4/0.6 \\ 
 & RFE (Hz/s) & \multicolumn{1}{c}{0.4/-/-} & 0.2/-/- & 0.013/0.013/0.024 & 0.006/0.006/0.006 & 0.301/0.240/0.322 & 0.008/0.006/0.006 \\
 & TVE (\%) & \multicolumn{1}{c}{1/-/-} & 1/-/- & 0.040/0.190/0.370 & 0.003/0.004/0.004 & 0.680/0.570/0.440 & 0.050/0.040/0.040 \\ \bottomrule
\end{tabular}}
\end{table*}

In AM and PM tests, the FBA-$H_2$ demonstrates clear improvements in all metrics, particularly with the larger modulation depths where both phasor techniques struggle. In PM tests, FBA-$H_2$ outperforms both alternatives, decreasing, for $a_{PM}=0.1$, the FE, RFE and TVE by 92\%/96\%, 64\%/84\%, and 85\%/0\% for IpDFT/CSTFM, respectively. Similarly, for $a_{PM}=0.5$, the error metrics for FE, RFE and TVE are reduced by 95\%/98\%, 90\%/95\% and 96\%/92\% for IpDFT/CSTFM, respectively. 
FBA-$H_1$ again proves to be insufficient in these tests, since the spectra of AMs and PMs have sidebands at $f_0\pm f_{AM}$ and $f_0\pm f_{PM}, \pm 2f_{PM}, ...$ \cite{AMPM}. Filter  $H_1$ attenuates the lower frequency sidebands for large modulations, introducing oscillations in the analytic signal components. Finally, in the FR tests, the instantaneous frequency increases from 45 Hz to 55 Hz at different rates. The FBA-$H_2$ performs equivalently or better in all cases.

In summary, while the performance of the phasor-based methods depends on the type and severity of the signal dynamic, the FBA-$H_2$ consistently delivers good results with relatively uniform errors across all tests. Moreover, $H_2$ significantly outperforms $H_1$, justifying the need for a high-order Hilbert filter to ensure a quality analytic signal. Consequently, $H_2$ is exclusively employed for all subsequent analysis

\subsubsection{Step transition tests}\label{sec:steptests}

FBA compliance with the standard step tests is verified using the shifted repeated signal method (SRS) in \cite{steptests_Standard} to calculate "equivalent samplings". The location of the step is shifted relative to the reporting time by $b/(10F_{RR})$ for $b=0...9$, and the estimates are interleaved to yield a higher resolution response curve. This allows improved estimates of the PMU Standard metrics: response time $R_T$, delay time $D_T$ and overshoot $OS$ \cite{PMUStandard,steptests_Standard}. The following tests use a 50 fps reporting rate for a resolution of 2 ms.

\begin{figure}[]
  \centering
\begin{tabular}{@{}cc@{}}
		\subfloat[Amplitude step test.]{\includegraphics[width=\columnwidth]{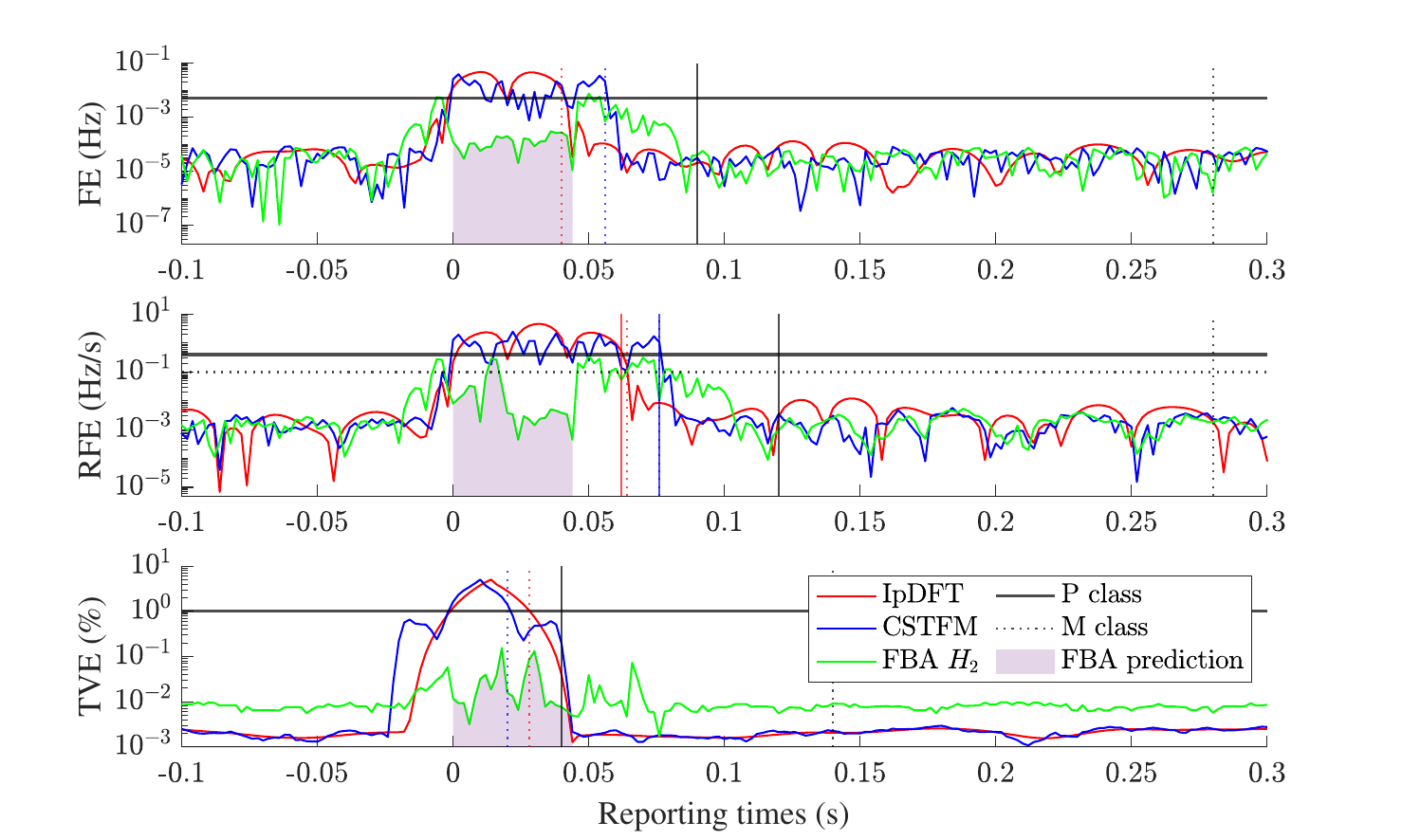}} \\ 
\subfloat[Phase step test.]{\includegraphics[width=\columnwidth]{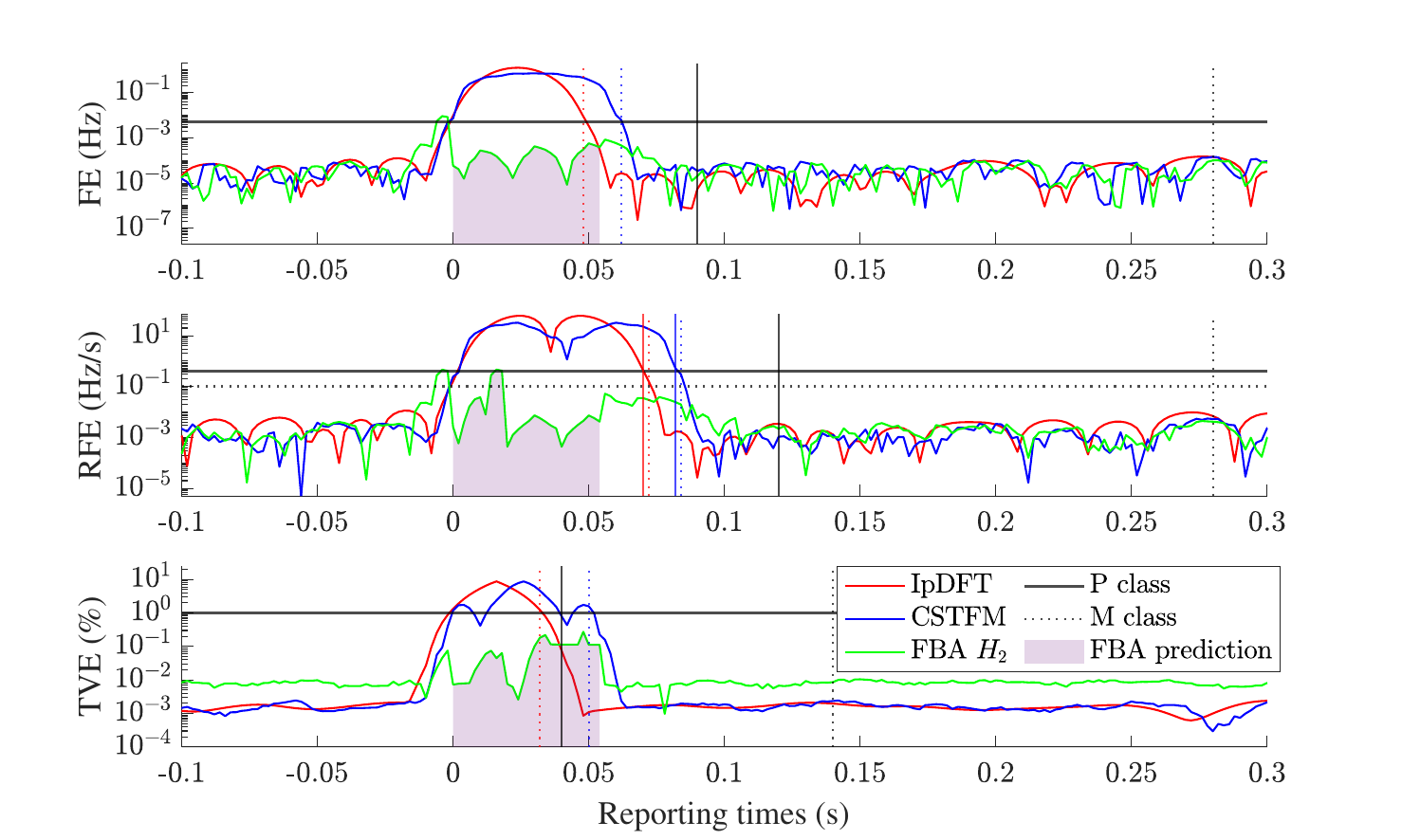}} &
\end{tabular}
\begin{tabular}{@{}cc@{}}
\subfloat[Measurement delay time for AS.]{\includegraphics[width=0.498\columnwidth]{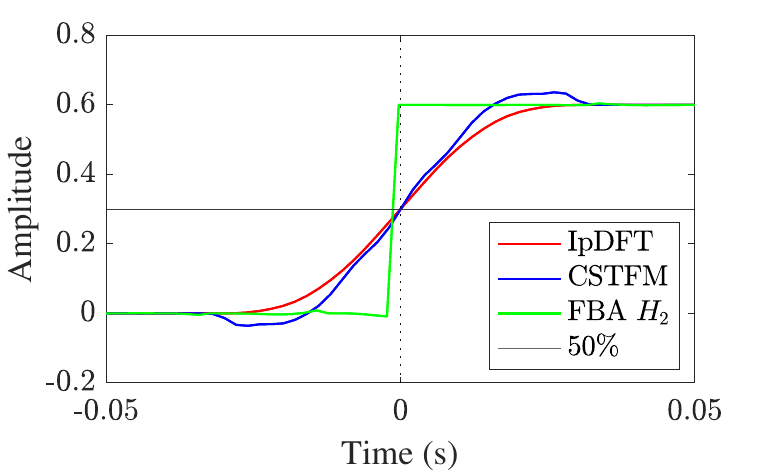}} & 
\subfloat[Measurement delay time for PS.]{\includegraphics[width=0.498\columnwidth]{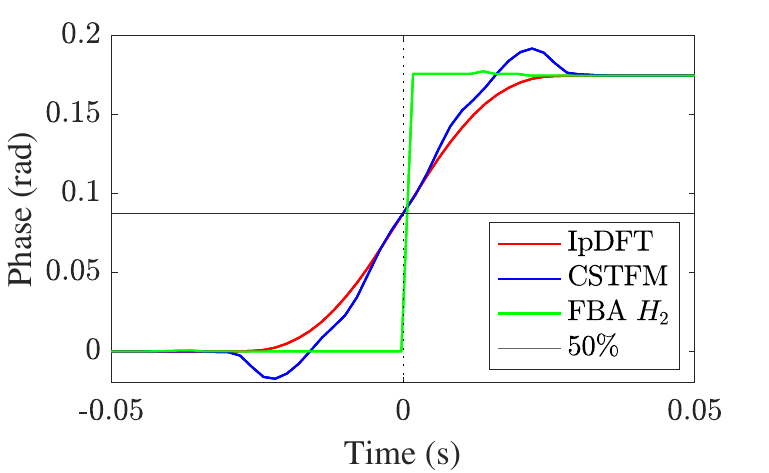}}
\end{tabular}
\caption[]{Performance comparison for AS (a and c) and PS (b and d) tests.}
\label{fig:AS_PS}
\end{figure}

Table \ref{tab_step} summarizes performance for AS and PS tests while Fig. \ref{fig:AS_PS} (a) and (b) display the interleaved error estimates, the P and M class thresholds and their corresponding response times, and the response time for each method, measurement, and threshold. Note that, for the phasor algorithms, the curves are aligned so that the measurement error that first surpasses the M class threshold is at $t=0$. However, since the FBA error does not or only briefly crosses the thresholds, performing the same shift would either not be possible, in the case of the TVE metric, or would yield a poor comparison. For example, for the AS test, the FBA reports over-threshold FE for a total of only 8 ms but $R_T=60$ ms. For RFE, $R_T=90$ ms for the M class threshold but 0 ms for the P class, and the TVE response time is 0 ms. For the PS tests, the response times are 4 ms, 24/22 ms, and 0 ms for FE, RFE (M/P class) and TVE, respectively. For this reason, $R_T$ for the FBA are not reported in Table \ref{tab_step} as they represent invalid comparisons in most cases. Consequently, in Fig. \ref{fig:AS_PS} (a) and (b), the first FBA estimate that detects the step is aligned with $t=0$. 

As expected, phasor-based methods perform poorly in the presence of a step and yield extreme frequency and ROCOF errors (e.g., IpDFT reports max FE $=1.2$ Hz for the PS test) while the errors for the FBA method barely exceed the thresholds (e.g., the max FE $=7.5$ mHz and RFE $=0.36$ Hz/s for the AS test). An explanation is needed to justify the results of these tests: the FBA  identifies steps and, while the step is within the observation window, offers interim state predictions based on prior estimates until a suitable post-event estimate is feasible. Consequently, because the step tests involve transitions between two identical SS conditions (i.e., 50 Hz to 50 Hz), the FBA yields low FE and RFE. If the post-step state differs significantly, the reference values would not match the FBA prediction and the errors would increase, as shown in the next section. To evaluate $D_T$, the stepped value at the center of the window (i.e., amplitude or phase) for each shifted waveform is recorded and interleaved, yielding a high resolution step response. The FBA excels in these tests as, by the time the step reaches the center of the observation window, it can provide accurate estimates of the post-step amplitude and phase and correct the model at the estimated step location (identified with an accuracy of $\pm2 T_s=\pm 0.2$ ms), resulting in the sharp transitions seen in Fig. \ref{fig:AS_PS} (c) and (d). For all methods, $D_T<2$ ms and are within the PMU Standard requirements.

More valuable indicators would be the time between the detection of the step and the availability of post-step estimates for phase, amplitude, frequency, and ROCOF. As shown by the predicted estimates (i.e., the purple shaded area) in Fig.
\ref{fig:AS_PS} (a) and (b), roughly 39 ms (AS) and $55$ ms (PS) elapse from when the step is detected in the window to when a new post-step estimates of the frequency and ROCOF are available. The running mean on the envelope reports 50\% of the stepped value after 3 ms and the full magnitude after $13.6$ ms, while the PS depth is approximated less than 10 ms after the step.  

\begin{table}[]
\caption{Amplitude and Phase Step Tests\label{tab_step}}
\renewcommand{\arraystretch}{1}
\centering
\begin{tabular}{ccccccc}
\toprule
 & \multicolumn{1}{c}{\textbf{Metric}} & \textbf{Error} & \textbf{Limit} & \textbf{IpDFT} & \textbf{CSTFM} & \textbf{FBA} \\ \midrule
 \parbox[t]{2mm}{\multirow{8}{*}{\rotatebox[origin=c]{90}{Amplitude Step}}} & \multicolumn{1}{c}{\multirow{3}{*}{Max}} & FE (mHz) & - & 47 & 39 & 7.5 \\  
 & \multicolumn{1}{c}{} & RFE (Hz/s) & - & 4.64 & 2.48 & 0.36 \\  
 & \multicolumn{1}{c}{} & TVE (\%) & - & 5 & 5.04 & 0.15 \\ \cmidrule{2-7} 
 & \multicolumn{1}{c}{\multirow{3}{*}{$R_T$ (ms)}} & FE (P/M) & 90/280 & 40 & 56 & - \\  
 & \multicolumn{1}{c}{} & RFE (P/M) & 120/280 & 60/64 & 76/76 & - \\  
 & \multicolumn{1}{c}{} & TVE (P/M) & 49/140 & 28 & 20 & - \\ \cmidrule{2-7} 
 & \multicolumn{1}{c}{$D_T$ (ms)}&  & 5 & \textless{}2 & \textless{}2 & \textless{}2 \\ \cmidrule{2-7} 
 & \multicolumn{1}{c}{OS   (\%)} & & 5/10 & -0.006 & 6 & 0.006 \\ \midrule
\parbox[t]{2mm}{\multirow{8}{*}{\rotatebox[origin=c]{90}{Phase Step}}}& \multicolumn{1}{c}{\multirow{3}{*}{Max}} & FE (mHz) & - & 1266 & 715 & 9.1 \\  
 & \multicolumn{1}{c}{} & RFE (Hz/s) & - & 60.10 & 32.22 & 0.46 \\  
 & \multicolumn{1}{c}{} & TVE (\%) & - & 8.7 & 8.7 & 0.27 \\ \cmidrule{2-7} 
 & \multicolumn{1}{c}{\multirow{3}{*}{$R_T$ (ms)}} & FE (P/M) & 90/280 & 48 & 62 & - \\  
 & \multicolumn{1}{c}{} & RFE (P/M) & 120/280 & 68/72 & 78/84 & - \\  
 & \multicolumn{1}{c}{} & TVE (P/M) & 49/140 & 32 & 50 & - \\ \cmidrule{2-7} 
 & \multicolumn{1}{c}{$D_T$ (ms)} & & 5 & \textless{}2 & \textless{}2 & \textless{}2 \\ \cmidrule{2-7} 
 & \multicolumn{1}{c}{OS   (\%)} & & 5/10 & -0.01 & 9.82 & 1.5 \\ \bottomrule
\end{tabular}
\end{table}

\subsection{Multidynamic Tests}\label{sec_multidyn}

While the FBA performs quite well on the standard dynamic tests, the test waveforms do not effectively showcase the algorithm's full capabilities. This section explores more appropriate and challenging exams to investigate the limits of the FBA, analyzing signals that that go beyond the PMU Standard and reporting performance indicators that reflect the dynamic nature of these signals. The two considered waveforms are:

\begin{itemize}
    \item \textbf{AM/PM}: $A_0=0.6$ p.u., $a_{AM}=10$ \%, $a_{PM}=0.1$ rad, $f_{AM}=f_{PM}=5$ Hz, $f_0=50$ Hz. As in \cite{PMUStandard}, PM and AM are 180 degrees out of phase.% to mimic two machines oscillating against each other.
   \item \textbf{AS/PS then FR}: $A_0=0.6$ p.u., $a_{AS}=10$ \%, $a_{PS}=~\pi/18$ rad, $f_0=50$ Hz,  $R=3$ Hz/s.% 
\end{itemize}

The following metrics are also reported:

\begin{itemize}
     \item The instantaneous RFE (IRFE) is the derivative of the instantaneous frequency at the reporting time, as opposed to the differential ROCOF in \ref{sec_single_dyn_test}, and is only available for the FBA method. 
     
     \item  The Time-Domain Error (TDE) reflects how closely the estimated model $\hat{x}(t_l)$ matches the input waveform $x(t_l)$. Note that only the non-overlapping portion of the sliding windows is analyzed. For a window length of 60 ms and $F_{RR}=50$ fps, the central 20 ms (200 samples) are used to calculate the TDE as follows:
\begin{equation}
    TDE=\sum_{l=200}^{L-201}|x(t_l)-\hat{x}(t_l)|
\end{equation}
\end{itemize}

\begin{figure}[]
  \centering
\begin{tabular}{@{}cc@{}}
		\subfloat[Modulation test: AM/PM]{\includegraphics[width=\columnwidth]{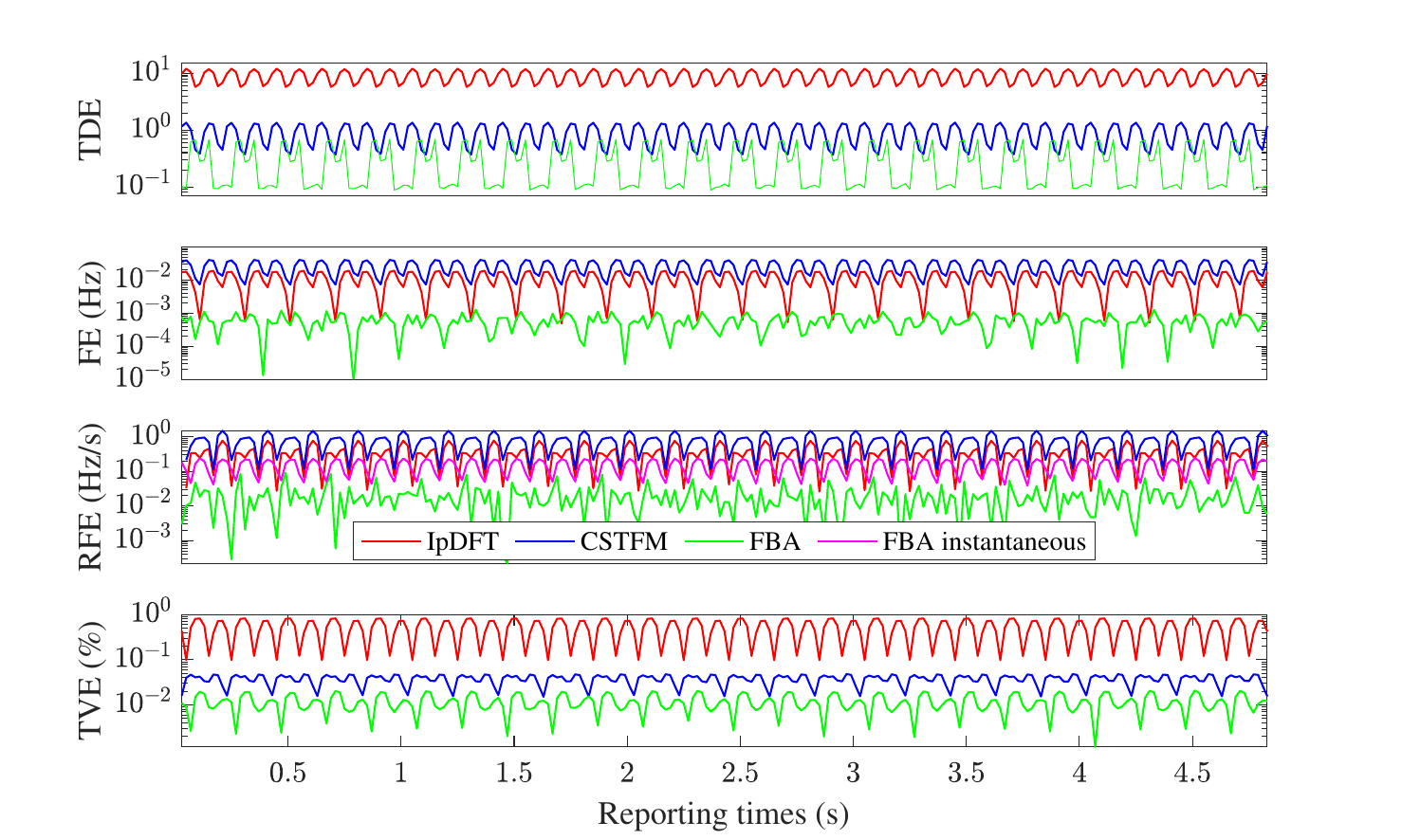}} \\ 
\subfloat[Step test: AS/PS followed by FR at 3 Hz/s.]{\includegraphics[width=\columnwidth]{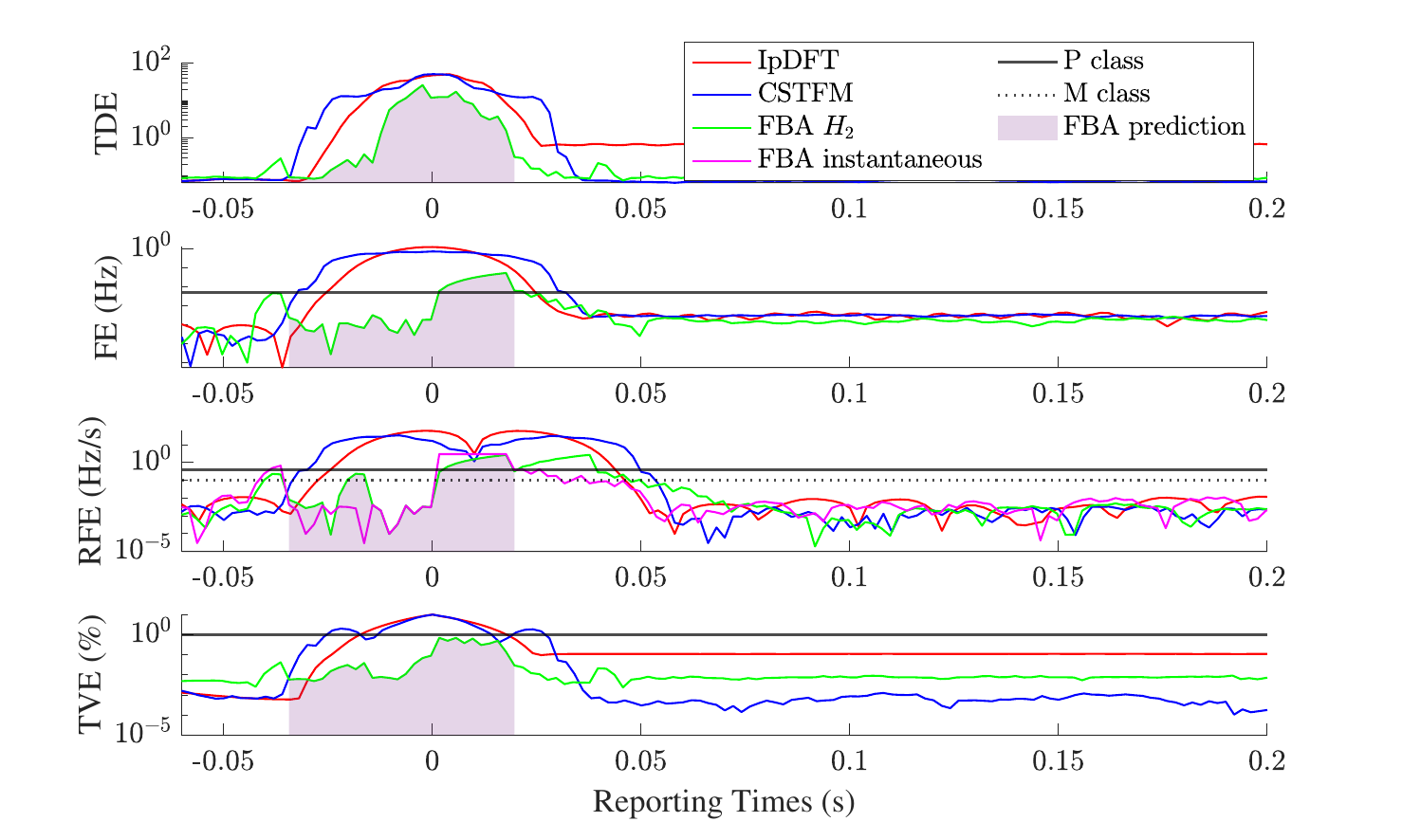}} &
\end{tabular}
\caption[]{Performance comparison for multidynamic tests.}
\label{fig:multidynamic}
\end{figure}

The results for the AM/PM test in Fig. \ref{fig:multidynamic}(a) show the FBA clearly outperforming the other methods in all metrics, with orders of magnitude improvement in some cases. The second test involves an AS/PS followed by a FR. Since the pre- and post-step state do not match, the FBA does not have an unfair advantage. Fig. \ref{fig:multidynamic}(b) shows all metrics, computed with the SRS technique. All reported errors are aligned such that, at $t=0$, the step is in the center of the window. Observe the shaded portion, where the FBA detects the step and begins predicting the state. When the step reaches the center of the window, the FBA errors increases as the post-step state does not matches the prediction. However, once a post-step estimate is available, the TDE, FE, and TVE immediately decrease. 

Furthermore, since the differential ROCOF requires two consecutive estimates of the frequency, there is significant delay before post-step ROCOF can be identified. The FBA, on the other hand, provides a below-threshold estimate of the frequency and ROCOF  6.3/12.3 ms and 24.3/30.0 ms before the IpDFT/CSTFM, respectively. Thus, demonstrating how the FBA can rapidly identify the post-step state. Next, observe that, as the step moves through the window, the TDE reported by the FBA is largely unaffected, since the pre-step state is preserved. The TDE increases once the step enters the central 20 ms of the window, but decreases as the FBA achieves better post-step estimates of the envelope, phase offset and, eventually, the frequency. Finally, despite the more challenging case, the TVE never exceeds the standard limit of 1\%.

%% file: Chapters/6_Conclusions.tex
\section{Conclusion}\label{sec:Conclusion}

This paper has presented a novel signal processing algorithm tailored for the identification and characterization of signal dynamics in power systems. The FBA method addresses several of the limitations inherent in state-of-the-art phasor-based techniques, including the simplified signal model and the periodicity assumptions inherent in Fourier analysis. 

The results of the validation tests showed the advantages of the FBA in identifying and characterizing signal dynamics like AM and PM, as compared to static and dynamic phasor methods. The PMU Standard step tests and metrics were found insufficient for evaluating the algorithm's performance, as the FBA can promptly detect the step and estimate the post-event state. Indeed, even for the AS/PS/FR signal, the reported metrics either did not or only briefly exceeded the standard thresholds. The FBA maintains the pre-step estimate and reports the post-step state less than 20 ms after the step passes the center of the window. Finally, since the FBA approximates the instantaneous frequency variation, it can provide post-step frequency and ROCOF estimates prior to phasor-based methods.

Thus, the FBA allows for the compression of complex signal dynamics into a parameterized model that better reconstructs measured non-stationary waveforms in modern power grids. Consequently, the FBA could be integrated into relays for protection schemes, flagging the presence of steps in order to avoid inappropriate Loss-of-Mains triggering \cite{Bornholm}, improve ROCOF estimations to better guide under-frequency load shedding (UFLS) decisions, or complement PMUs, providing operators with nuanced information on the state of the grid.

%% file: Chapters/Appendix.tex
\section*{Appendix}\label{p1_appendix}
\subsection*{Cramer-Rao Lower Bound (CRLB) }

To derive the CRLB, a theoretical bound on the variance of an estimator, for FR estimation, the analytic signal argument is: $x_\phi(t_l)=x_{FR}(t_l,\boldsymbol{\beta})+\varepsilon_l$ where $\varepsilon$ is assumed to be additive Gaussian noise $\mathcal{N}(0,\sigma^2)$ with a standard deviation of $\sigma$. First, the joint probability density function is defined as:
\begin{align}
    pdf&(x_\phi(t_l), \boldsymbol{\beta})=
    \prod_{l=0}^{L-1} \frac{1}{\sqrt{2\pi\sigma^2}}e^{-\frac{1}{2\sigma^2}(({x_\phi(t_l)-x_{FR}(t_l,\boldsymbol{\beta}))})^2} \nonumber \\
    &=\frac{1}{(2\pi\sigma^2)^{L/2}}e^{-\frac{1}{2\sigma^2}\sum_{l=0}^{L-1}(x_\phi(t_l)-x_{FR}(t_l,\boldsymbol{\beta}))^2}
\end{align}

With the log likelihood function $LL=\ln(pdf)$, the Fisher information, $\mathcal{I}_{\beta_i}=\mathbb{E}[-\partial^2/\partial\beta_i^2 [LL(\boldsymbol{\beta}|x_\phi(t_l))]]$, and the $CRLB_{\beta_i}=1/\mathcal{I}_{\beta_i}$, the following can be derived:
\begin{equation}
    CRLB_{\beta_0}=\sigma^2/L,
\end{equation}
\begin{equation}
    CRLB_{\beta_1}=
    6\sigma^2/(T_s^2 L(L-1)(2L-1)),
\end{equation}
\begin{equation}
    CRLB_{\beta_2}=30\sigma^2/(T_s^4 L(L-1)(2L-1)(3L^2-3L-1))
\end{equation}

The uncertainty in the estimation of $\beta_0$, $\beta_1$ and $\beta_2$ is then inversely proportional to $L$, $L^3$ and $L^5$, respectively.